\documentclass[pdflatex,sn-basic]{sn-jnl}

\usepackage{graphicx}
\usepackage{amsmath,amssymb,amsfonts}
\usepackage{textcomp}
\usepackage{multirow}
\usepackage[table]{xcolor}
\usepackage{array}
\usepackage{makecell}
\usepackage{enumitem}
\usepackage{listings}
\usepackage{algorithmic}

\begin{document}

\title[Code Smells in Efferent Neighbors and Class Stability]
{The Influence of Code Smells in Efferent Neighbors on Class Stability}

\author*[1]{\fnm{Zushuai} \sur{Zhang}}
\email{derek.zhang@auckland.ac.nz}

\author[1]{\fnm{Elliott} \sur{Wen}}
\email{elliott.wen@auckland.ac.nz}

\author[1]{\fnm{Ewan} \sur{Tempero}}
\email{e.tempero@auckland.ac.nz}

\affil*[1]{%
  \orgname{University of Auckland},
  \orgaddress{
    \city{Auckland},
    \country{New Zealand}
  }
}

\abstract{%

Understanding what drives code instability is essential for effective software maintenance, as unstable classes require larger or more frequent edits and increase the risk of unintended side effects. Although code smells are widely believed to harm maintainability, most prior stability studies examine only the smells within the class being modified. In practice, however, a class can change because its efferent neighbors (i.e., the classes it depends on) are modified due to ripple effects that propagate along static dependencies, even if the class itself is clean. Such ripple effects may be more severe when efferent neighbors exhibit code smells. Therefore, this study investigates whether code smells in a class's efferent neighbors affect its stability. We mined one year of commit history from 152 top-starred GitHub Java projects, detected code smells and static dependencies, and modeled their associations with class stability. Our results challenge the practice of examining code smells only within the class being maintained by showing that this provides an incomplete view of stability risk. Specifically, considering code smells in efferent neighbors reveals stability risks that are not observed when examining code smells in the class itself. Moreover, considering smells in both the class and its efferent neighbors, as well as their relationships across dependencies, reveals further stability risks that are not observed when smells at these two locations are analyzed separately. Consequently, these findings suggest that code smell analysis for class maintenance should extend beyond the class being maintained to the classes on which it depends.

}

\keywords{%
code smells, software evolution, software maintenance, dependency analysis, empirical software engineering
}

\maketitle

\section{Introduction}\label{sec:intro}

Changing source code is the heartbeat of software maintenance. For most maintenance tasks, such as fixing bugs, refactoring, or delivering new features, teams aim to keep edits small and localized because smaller changes cost less, disturb less context, and reduce the risk of unintended side effects \citep{kamei2012large}. A system exhibits low stability when parts of it require large edits during maintenance tasks or are frequently modified across multiple tasks \citep{olbrich2009evolution, santana2024unraveling}. Such instability greatly increases the effort and risk of making changes. It is therefore important to understand which code characteristics drive this instability, enabling teams to address these factors, maintain system stability, and deliver changes effectively and efficiently.

\begin{figure}[h]
\centering
  \includegraphics[width=5cm]{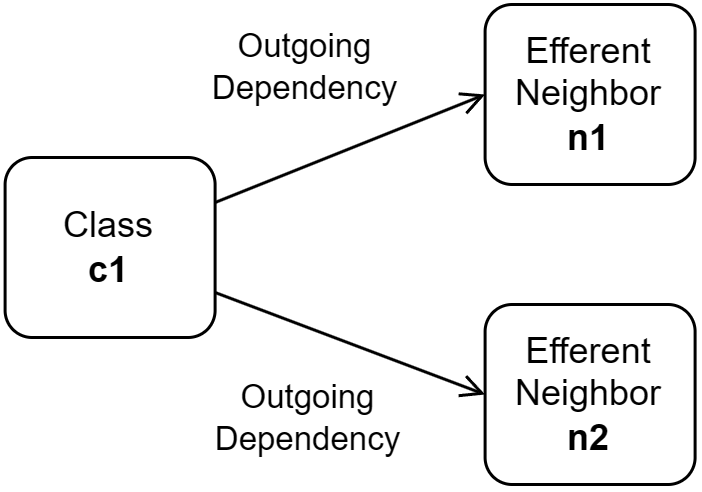}
  \caption{Example of efferent neighbors: class \texttt{c1} depends on \texttt{n1} and \texttt{n2} via outgoing dependencies.}~\label{fig:efferent-concept}
\end{figure}

One long-suspected factor is the presence of code smells: surface-level characteristics in source code that may indicate underlying design problems \citep{fowler1999refactoring}. Code smells are widely believed to degrade maintainability \citep{fowler1999refactoring, olbrich2010all, ban2014recognizing}. Prior studies report that smelly classes tend to be edited more frequently and with greater change size than their non-smelly peers because of their large, complex, and non-cohesive nature \citep{santana2024unraveling, khomh2009exploratory, palomba2018diffuseness, olbrich2009evolution}. However, when developers change a class, it is not always because the class itself requires modification. A class may change because the classes it depends on via outgoing dependencies have changed, which can create ripple effects along static dependencies, even if the class itself is not smelly \citep{salama2019stability}. We refer to these dependent classes as the class's \emph{\textbf{efferent neighbors}}\footnote{Here, \emph{efferent} is used as an adjective meaning ``conducting outward'' (i.e., outgoing), to emphasize that these neighbors are reached via outgoing dependencies from the class. Definition source: \url{https://www.merriam-webster.com/dictionary/efferent}}. Figure~\ref{fig:efferent-concept} illustrates the concept: class \texttt{c1} has outgoing dependencies on classes \texttt{n1} and \texttt{n2}, which we treat as \texttt{c1}'s efferent neighbors. Changes in \texttt{n1} or \texttt{n2} can propagate to \texttt{c1} through these dependencies, creating ripple effects that may require \texttt{c1} to change as well. Consequently, when the efferent neighbors are smelly, these ripple effects may occur more frequently, thereby reducing the class's stability. Focusing only on smells within the class captures the first situation but overlooks dependency-driven changes. To assess whether code smells also influence the stability of a class through this second pathway, we must analyze the smells present in its efferent neighbors and explicitly account for the static dependencies between classes.

In addition, prior work suggests that code smells rarely occur in isolation. Multiple smells frequently appear together within the same class or in classes connected by static dependencies \citep{palomba2018large, yamashita2015inter}. This phenomenon is referred to as \emph{\textbf{code smell interrelation}}, and multiple code smells can collectively exacerbate maintainability issues \citep{abbes2011empirical, yamashita2013exploring, yamashita2015inter}. A further consequence of this interrelation is \emph{\textbf{code smell interaction}}, in which two smells are directly connected through static dependencies. These interactions may further complicate maintainability by affecting modularity, comprehensibility, and modifiability because of the direct structural connection between smelly artifacts \citep{yamashita2013exploring, zhang2025analyzing}. However, to the best of our knowledge, few studies have empirically examined how code smell interrelation and interaction influence code quality. Therefore, the objective of this study is to assess \emph{the association between code smells in a class's efferent neighbors and its stability, while also considering code smell interrelation and code smell interaction between smells in the class and its efferent neighbors.}

For this purpose, we mined one year of commit history from 152 top-starred open-source Java projects. We detected 10 code smells, including 5 at the class level and 5 at the method level, and extracted static dependencies to identify code smell interrelation and interaction. We analyzed each commit to obtain class-level stability measurements and modeled these data using negative binomial generalized linear mixed models with project-level random intercepts and appropriate controls.

Our contributions are as follows: \textbf{(i)} the design and execution of a large-scale empirical study, with the source code, analysis scripts, and dataset made available through our replication package \citep{effint2026zhang} to support replication; and \textbf{(ii)} the first rigorous empirical evidence on the influence of code smells in efferent neighbors on class stability, providing empirically grounded insights that can inform maintenance prioritization and guide future research on smells and their relationships across dependencies.

\section{Background and Related Work} \label{sec:background}

\subsection{Code Smells} \label{subsec:ten_code_smells}

Code smells are surface-level characteristics in source code that can indicate deeper design issues \citep{fowler1999refactoring}. There are different types of code smells, and they can be classified into class-level smells and method-level smells, as some affect the quality of a single method while others impact an entire class \citep{lanza2007object}. In this study, we focus on 10 types of code smells \citep{lanza2007object}. The method-level smells are \emph{Feature Envy} (FE), \emph{Brain Method} (BM), \emph{Dispersed Coupling} (DP), \emph{Intensive Coupling} (IT), and \emph{Shotgun Surgery} (SG). The class-level smells are \emph{God Class} (GC), \emph{Brain Class} (BC), \emph{Data Class} (DC), \emph{Refused Bequest} (RB), and \emph{Tradition Breaker} (TB). Table~\ref{tab:code_smells} details their definitions \citep{lanza2007object}.

\begin{table*}[t]
\centering
\footnotesize
\caption{Definitions of code smells considered in this study \citep{lanza2007object}}
\renewcommand{\arraystretch}{1.4} %
\setlength{\tabcolsep}{6pt}
    \begin{tabular}{p{0.27\textwidth} p{0.66\textwidth}}
    \hline
    \textbf{Code smell} & \textbf{Definition} \\
    \hline
    
    \textbf{Feature Envy (FE)} &
    A method that uses another class's data more than the data of its own class. \\
    \hline
    
    \textbf{Brain Method (BM)} &
    A method that is excessively large and complex, uses many variables, and is deeply nested, tending to centralize too much functionality within a single class. \\
    \hline

    \textbf{Dispersed Coupling (DP)} &
    A method that interacts with an excessive number of different classes by invoking only one or a few methods from each. \\
    \hline

    \textbf{Intensive Coupling (IT)} &
    A method that invokes an excessive number of methods from one or a few other classes. \\
    \hline

    \textbf{Shotgun Surgery (SG)} &
    A method that is invoked by too many other methods from many different classes. \\
    \hline
    
    \textbf{God Class (GC)} &
    A large and complex class that concentrates much of the system's functionality, frequently uses data from other classes, and has low cohesion, suggesting multiple responsibilities. \\
    \hline
    
    \textbf{Brain Class (BC)} &
    A large and complex class that accumulates too much of the system's functionality and contains at least one Brain Method. It exhibits low cohesion, but can be more cohesive than a God Class. \\
    \hline
    
    \textbf{Data Class (DC)} &
    A low-complexity class that mainly stores data and provides access through public fields or accessor methods rather than encapsulating behavior. \\
    \hline

    \textbf{Refused Bequest (RB)} &
    A subclass that makes little use of inherited protected members and rarely overrides parent methods. \\
    \hline

    \textbf{Tradition Breaker (TB)} &
    A subclass that adds many new public methods while rarely specializing inherited ones, making its interface largely unrelated to its superclass. \\
    \hline
    
    \end{tabular}
    \label{tab:code_smells}
\end{table*}

Although artifacts may exhibit additional code smells beyond the 10 we consider, our selection follows established practice in software engineering research. We focus on a subset of smells that (i) offer balanced coverage across both levels, (ii) include smells that are widely studied in software maintenance research, occur frequently in codebases, and have been linked to reduced maintainability \citep{li2007empirical, cairo2018impact, tufano2017and, palomba2018diffuseness, kaur2020systematic, reis2020code, zakeri2023systematic}, (iii) have clear definitions and widely accepted detection techniques \citep{lanza2007object}, (iv) are supported by validated detection tools \citep{fernandes2016review}, and (v) align with prior studies, enabling meaningful comparison \citep{santana2024unraveling}. Our goal is not to exhaustively cover all possible smells, but to analyze a representative subset. Consequently, we believe this choice makes our results more interpretable and practical.

We do not include a few well-known code smells such as \emph{Long Method} (i.e., a method that is too long \citep{fowler1999refactoring}) and \emph{Large Class} (i.e., a class that is too big \citep{fowler1999refactoring}) in this study. Although widely discussed, these smells are usually detected using a single aspect, such as method or class length, which on its own is often an overly sensitive heuristic (e.g., a \emph{Large Class} that is fully cohesive and has a clear single responsibility, or a \emph{Long Method} that is not deeply nested, accesses only a limited number of variables, and is easy to understand). In contrast, the smells we analyze are defined based on multiple aspects (e.g., the definitions of \emph{Brain Method} and \emph{God Class} include length and complexity, but also consider other aspects such as cohesion and nesting level), using richer structural and coupling-related characteristics that are more directly aligned with our dependency-based theory. Therefore, we focus on the 10 well-defined smells to strengthen construct validity by aligning smell definitions with our dependency-based theory, while reducing detection noise from single-aspect, overly sensitive heuristics.

\subsection{The Influence of Code Smells on Code Stability}

Software stability is a sub-attribute of maintainability \citep{iso9126}. Prior research has examined stability from different perspectives, each focusing on a particular category of software artifact. \citet{salama2019stability} classified these perspectives into four levels, where a level refers to the category of artifact being analyzed. The four levels are the code level, the architecture level, the design level, and the requirements level. In this paper, we focus on the code level, as this is where code smells manifest; specifically, we are interested in the stability of class artifacts. At this level, stability is defined as \citep{salama2019stability}: (i) \emph{\textbf{intrinsic stability}}: the ability to remain largely unchanged over time (this captures an intrinsic property of the artifact itself, independent of external pressures such as ripple effects); and (ii) \emph{\textbf{ripple-effect stability}}: the ability to resist the ripple effect of changes (an example of low ripple-effect stability is when one class changes and the classes depending on it also require modification).

Different metrics exist to measure code stability. \citet{martin1994oo, martin2000design} proposed a package instability metric, which is proportional to a package's efferent coupling, that is, the number of classes outside the package on which the classes inside the package depend. The rationale is that a package whose classes rely on many external classes is more likely to change when any of those dependent classes change. This conceptualization can be extended to the class level using the same rationale, where the \emph{\textbf{efferent coupling}} of a class corresponds to the number of other classes it depends on (i.e., its efferent neighbors). In software evolution research, code stability is typically measured using two process metrics~\citep{olbrich2009evolution, khomh2009exploratory, santana2024unraveling}: \emph{change frequency}, which measures how often a software entity is modified within a specific period, and \emph{change size}, which quantifies the magnitude of each change and is usually expressed in terms of lines of code added or deleted (code churn) during each commit.

Empirical studies generally report that code smells are associated with lower code stability. Several investigations show that smelly classes tend to change more frequently and often with larger modifications, with this effect becoming stronger when multiple smells occur in the same class \citep{santana2024unraveling, khomh2009exploratory, palomba2018diffuseness, olbrich2009evolution}. Similar patterns were reported in server-side web applications, where smelly files were more frequently updated and exhibited greater change size \citep{bessghaier2021longitudinal}. A notable exception was reported by \citet{olbrich2010all}, who found that although God and Brain Classes exhibited higher change frequency and more defects in absolute terms, these differences disappeared or reversed once class size was controlled for, indicating that size is a confounding variable in observed stability differences.

\subsection{Architectural Smells and Their Evolution}
\label{subsec:arch_smell_evol}

Architectural smells are architecture-level design decisions or structural patterns that negatively impact system quality attributes \citep{garcia2009toward}. The key difference between architectural smells and code smells lies in the artifacts in which they manifest. Code smells are defined in terms of code-level (i.e., implementation-level) artifacts, such as methods and classes, whereas architectural smells are defined in terms of architectural-level artifacts, such as components and subsystems \citep{garcia2009toward}. For example, the code smell \emph{God Class} and the architectural smell \emph{God Component} \citep{azadi2019architectural} both describe an artifact that has too many responsibilities. However, \emph{God Class} refers to an overloaded class, while \emph{God Component} refers to an overloaded component.

Previous research has studied the evolution of architectural smells. \citet{sas2019investigating} tracked how individual smell instances evolve over time by mining three smell types across 14 open-source Java projects. They showed that smell types differ in multiple aspects, such as growth rate and lifetime, offering guidance for smell refactoring and prioritization. Extending this work, \citet{sas2022evolution} analyzed how smells evolve in industrial systems. They tracked smells in nine C/C++ projects and complemented mining with interviews with 12 developers. Their findings suggest that smells can persist across releases, overlap within the same components, and be perceived by developers as contributing to maintainability difficulties. Developers also described how smell analysis can provide actionable refactoring insights.

\citet{gnoyke2021evolutionary} studied the evolution of smells and their impact on system degradation by analyzing 14 open-source Java systems. They reported that smells remain mostly stable relative to code size, although some smell types have a greater impact on system degradation. They also identified properties that influence the strength of smells' impact on system degradation. Recently, \citet{gnoyke2024evolution} expanded this work by characterizing typical evolution patterns of smells. They found that some smells tend to accumulate and merge into more complex structures over time, while others remain relatively stable.

In this study, we focus on code smells because our goal is to analyze the stability of classes, which are code-level artifacts rather than architectural-level artifacts.

\subsection{The Influence of Interrelated and Interacting Code Smells on Maintainability}
\label{sec:int_code_smells_maintainability}

Code smells are widely recognized and have been empirically shown to undermine maintainability \citep{fowler1999refactoring, olbrich2010all, ban2014recognizing}. However, they rarely occur in isolation. Instead, multiple code smells often appear together in the same class or in related classes, collectively worsening maintainability. This phenomenon is known as \emph{\textbf{code smell interrelation}} \citep{yamashita2013exploring}. \citet{yamashita2013exploring} categorized interrelations as follows: (i) \emph{\textbf{code smell collocation}}: multiple smells within the same class; and (ii) \emph{\textbf{code smell coupling}}: smells appearing in different classes that are connected through static dependencies.

Several studies have examined how the collocation of multiple smells within the same class affects maintainability. \citet{abbes2011empirical} found that individual smells showed little measurable impact on comprehension, whereas their collocation significantly increased the time, effort, and errors involved in understanding the code. Later work \citep{martins2021code, santana2024exploratory} showed that classes with multiple collocated smells tend to have higher complexity and lower cohesion, with some combinations, such as a God Class collocated with Feature Envy, being especially detrimental. Developers also reported that collocated smells make code harder to understand and modify \citep{martins2021code}.

Interestingly, prior work suggests that maintenance problems may arise not only from the presence of multiple interrelated smells, but from their interaction through static dependencies \citep{yamashita2012assessing, yamashita2015inter}. In an exploratory study, \citet{yamashita2013exploring} observed professional developers performing maintenance tasks and collected their reflections through interviews and think-aloud sessions. Developers indicated that collocated and coupled smells can interact along static dependencies, making code harder to follow and causing ripple effects during changes, which increase effort and defect risk. To the best of our knowledge, this remains the only study that directly links such interactions to maintainability, while later work has mainly identified recurring patterns of collocation and coupling without assessing their impact \citep{yamashita2015inter, fontana2015towards, walter2018code, sobrinho2021interplay, martins2020code}.

Recently, \citet{zhang2025analyzing} formally defined the term \emph{\textbf{code smell interaction}}: a code smell instance interacts with another code smell instance if there is a static dependency that directly connects the two instances. They concluded that such interactions are associated with degraded modularity.

\begin{figure}
\centering
  \includegraphics[width=8cm]{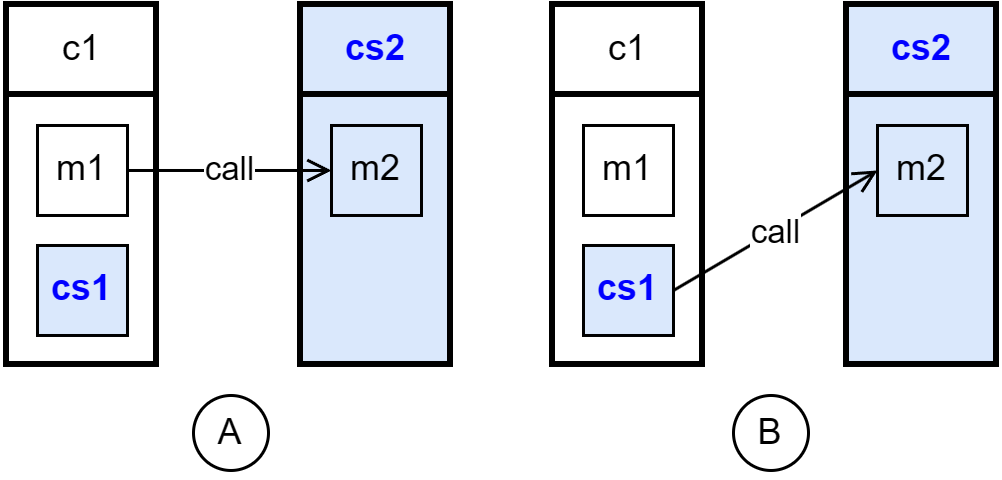}
  \caption{Examples of code smell coupling and interaction between a method-level code smell instance \texttt{cs1} in class \texttt{c1} and a class-level code smell instance \texttt{cs2}: (A) coupled but non-interacting code smells; (B) coupled and interacting code smells.}~\label{fig:coupled_interaction}
\end{figure}

The core difference between code smell coupling and code smell interaction is that coupling concerns dependencies between classes, while interaction concerns dependencies between individual code smell instances. Code smell instances can be more fine-grained than classes because some smells appear at the method level. For example, in Figure~\ref{fig:coupled_interaction}, class \texttt{c1} contains method \texttt{m1} and method \texttt{cs1}, where \texttt{cs1} is a method-level code smell instance. Class \texttt{cs2} contains method \texttt{m2}, where \texttt{cs2} is a class-level code smell instance. The two classes are connected by a method-call dependency in both cases \texttt{A} and \texttt{B}, so code smell coupling exists in both cases. However, code smell interaction exists only in case \texttt{B}, because the call dependency directly connects method \texttt{cs1} with class \texttt{cs2}.

In summary, while prior studies have extensively examined the influence of collocated code smells on maintainability, few have addressed coupled code smells or the broader factor of interaction. This gap is particularly relevant to code stability. Focusing solely on how code smells within a class affect its stability addresses only intrinsic stability, while overlooking ripple-effect stability. When efferent neighbors contain code smells, they may become unstable. As a result, ripple effects along static dependencies may be triggered more frequently, requiring the assessed class to change more often as well. To investigate this, our study incorporates code smells in efferent neighbors and accounts for both code smell interrelation and interaction.

\section{Study Design} \label{sec:study_design}

In this section, we outline the key components of our study. We begin by explaining the types of dependencies we consider and defining efferent code smell coupling and interaction. We then present our research questions and hypotheses, followed by the variable definitions and the procedures used to obtain and analyze them.

In this Java-based study, we operationalize a \emph{class} as either a Java class or interface. We include interfaces because they can exhibit code smells and participate in the static dependencies analyzed in this study, and can therefore be included among the classes whose stability is analyzed or occur as their efferent neighbors. We use the term \emph{class} in this broader sense throughout the paper, except in the dependency definitions in Section~\ref{sec:dep_definition}, where classes and interfaces are distinguished as separate artifact types.

\subsection{Dependency}\label{sec:dep_definition}

We utilize the following definition of dependency from \citet{stevens1974structured}: “...A dependency is when the functioning of one element A requires the presence of another element B...” We interpret this as a directional connection between two code artifacts, flowing from the source artifact to the target artifact. We capture the requirement between the two through different relations. For example, a \emph{method} (source artifact type) may \emph{create} (relation type) an object of a \emph{class} (target artifact type). We therefore define a dependency type as a triple:
\[
(\text{relation type},\ \text{source artifact type},\ \text{target artifact type}).
\]

\noindent
In this paper, we consider 5 artifact types: class, interface, field, constructor, and method. Following previous work \citep{yamashita2015inter, zhang2025analyzing}, we focus on 10 relation types: call, create, contain, cast, use, throws, return, parameter, extend, and implement. Table~\ref{tab:dependency_type} lists the valid source and target artifact types as sets for each relation. Every ordered pair formed by selecting one source artifact type and one target artifact type is valid.

\begin{table*}[tb]
    \centering
    \caption{Dependency types}
    \renewcommand{\arraystretch}{1.4} %
    \footnotesize
    \begin{tabular}{@{}p{1.15cm} p{3.0cm} p{2.9cm} p{4.7cm}@{}}
        \hline
        \textbf{Relation} & \textbf{Source Artifact} & \textbf{Target Artifact} & \textbf{Description} \\
        \hline
        call & \{Method, Constructor, Field\} & \{Method, Constructor\} & A method, constructor, or field initializer invokes a method or constructor. \\
        \hline
        create & \{Method, Constructor, Field\} & \{Class\} & A method, constructor, or field initializer instantiates (i.e., uses \texttt{new}) a class. \\
        \hline
        contain & \{Method, Constructor, Field\} & \{Class, Interface\} & A method or constructor declares a local variable of a class or interface type, or a field is declared with a class or interface type. \\
        \hline
        cast & \{Method, Constructor, Field\} & \{Class, Interface\} & A method, constructor, or field initializer explicitly casts an object to a class or interface type. \\
        \hline
        use & \{Method, Constructor, Field\} & \{Field\} & A method, constructor, or field initializer references a field. \\
        \hline
        throws & \{Method, Constructor\} & \{Class\} & A method or constructor declares that it may throw an exception (a class) or explicitly throws one within its body. \\
        \hline
        return & \{Method\} & \{Class, Interface\} & A method returns an object of a class or interface type. \\
        \hline
        parameter & \{Method, Constructor\} & \{Class, Interface\} & A method or constructor has a parameter of a class or interface type. \\
        \hline
        extend\textsuperscript{*} & \{Class, Interface\} & \{Class, Interface\} & A class extends another class, or an interface extends another interface. \\
        \hline
        implement & \{Class\} & \{Interface\} & A class implements an interface. \\
        \hline
        \noalign{\vskip 0.5em}
        \multicolumn{4}{@{}p{\linewidth}@{}}{%
            \textsuperscript{*}For \emph{extend}, only Class $\rightarrow$ Class and Interface $\rightarrow$ Interface are valid.
        } \\
    \end{tabular}
    \label{tab:dependency_type}
\end{table*}

We treat all considered dependency types uniformly when deriving dependency-based measurements and conducting analyses. Specifically, dependencies of different types contribute equally; we do not apply dependency-type-specific weights. We further discuss the implications of this decision in Section~\ref{sec:threats}.

\subsection{Efferent Code Smell Coupling and Interaction}\label{sec:eff_coup_int}

We use the term \emph{\textbf{focal class}} to refer to the class whose stability we are measuring, and the term \emph{\textbf{efferent neighbors}} to refer to all the classes it depends on.

If both the focal class and at least one of its efferent neighbors exhibit code smells, we refer to this as \emph{\textbf{efferent code smell coupling (EffCoup)}} \footnote{Here, \emph{efferent} modifies \emph{code smell coupling} as defined in Section~\ref{sec:int_code_smells_maintainability}, indicating that there exists an outgoing dependency that leads to code smell coupling}.

We employ the term \emph{\textbf{efferent code smell interaction (EffInt)}} \footnote{Here, \emph{efferent} modifies \emph{code smell interaction} as defined in Section~\ref{sec:int_code_smells_maintainability}, indicating that there exists an outgoing dependency that leads to code smell interaction} to refer to a situation in which a code smell instance $\mathrm{cs}_1$ of type $\mathrm{CS}_1$ interacts with a code smell instance $\mathrm{cs}_2$ of type $\mathrm{CS}_2$, where $\mathrm{cs}_1$ resides in the focal class, $\mathrm{cs}_2$ resides in its efferent neighbor, and there exists at least one outgoing dependency instance directly connecting $\mathrm{cs}_1$ to $\mathrm{cs}_2$, with $\mathrm{cs}_1$ as the source artifact and $\mathrm{cs}_2$ as the target artifact. We refer to such a dependency instance as an \emph{\textbf{efferent code smell interaction dependency instance (EffInt dependency instance)}}. We refer to $(\mathrm{cs}_1, \mathrm{cs}_2)$ as an \emph{\textbf{efferent code smell interaction smell-instance pair (EffInt smell-instance pair)}}. Their smell types form a directional \emph{\textbf{smell-type pair}}, denoted by $\mathrm{CS}_1\rightarrow\mathrm{CS}_2$. We refer to $\mathrm{CS}_1$ as the source smell type and $\mathrm{CS}_2$ as the target smell type. For a focal class, we define \emph{\textbf{the number of efferent code smell interactions (\texttt{\#EffInt})}} as the number of distinct EffInt smell-instance pairs, and we define \emph{\textbf{efferent code smell interaction intensity (\texttt{EffIntInten})}} as the total number of EffInt dependency instances connecting these pairs.

\subsection{Research Questions and Hypotheses}
\label{sec:rq_hypo}

In this section, we list our research questions (RQ) and formalize the hypotheses for each RQ. In addition to the primary analyses, we introduce exploratory research questions (ERQs) to obtain more fine-grained insights into how specific code smell types and smell-type pairs are associated with class stability. For all hypotheses, the null hypothesis ($H_0$) is that there is no statistically significant effect or association between the independent variable and the stability measure. Below, we list only the corresponding alternative hypotheses.

Additionally, stability was assessed based on change frequency and change size (see Section~\ref{sec:dep_variable}). Due to space constraints, each hypothesis is worded in terms of stability in general, but in the study, each one was evaluated as two separate hypotheses.

When we test the effect of efferent code smell coupling and interaction (RQ3 and RQ4), we examine both (a) the effect without accounting for the presence or number of code smells in the focal class and its efferent neighbors, and (b) the effect with these factors held constant. We do this because (i) deconfounding is necessary, since code smells in the focal class or its neighbors may independently decrease stability, and (ii) we want to determine whether coupling or interaction adds additional instability beyond what can be explained by the smells already present in these two locations (i.e., the focal class and its efferent neighbors). This allows us to distinguish the raw association from the unique contribution of coupling or interaction after controlling for baseline smell conditions.

\subsubsection{RQ1}

Although the influence of collocated code smells on stability has been studied extensively, we include this aspect for completeness and to enable comparison with previous literature. We consider not only whether code smells are present in the focal class, but also the number and variety of code smells it contains, because higher values of these measures are often associated with degraded maintainability, particularly reduced stability \citep{palomba2018diffuseness, santana2024unraveling}. Thus, our RQ1 and its hypotheses are:

\vspace{1mm}
\noindent
\fbox{%
  \parbox{\columnwidth}{%
    \hspace{3pt}%
    \parbox{0.95\columnwidth}{%
      \vspace{1mm}
      \textbf{RQ1:}
        Do code smells in a class influence its stability?
      \vspace{1mm}
    }%
    \hspace{3pt}%
  }%
}

\vspace{3mm}

\noindent
$\boldsymbol{H1.1}$: \textbf{Smelly classes} are significantly \textbf{less stable} than non-smelly classes.

\noindent
$\boldsymbol{H1.2}$: There is a significant \textbf{negative association} between the \textbf{number of code smells} a class exhibits and its stability.

\noindent
$\boldsymbol{H1.3}$: There is a significant \textbf{negative association} between the \textbf{variety of code smells} a class exhibits and its stability.

\medskip

RQ1 aggregates the measures across all smell types. To explore which specific code smell types are associated with stability, we further examine $\boldsymbol{H1.1}$ and $\boldsymbol{H1.2}$ for each code smell type. We do not consider $\boldsymbol{H1.3}$ because the variety of a single selected smell type reduces to whether that smell type is present, which is already addressed by $\boldsymbol{H1.1}$. Thus, our ERQ1 and its exploratory hypotheses are:

\vspace{1mm}
\noindent
\fbox{%
  \parbox{\columnwidth}{%
    \hspace{3pt}%
    \parbox{0.95\columnwidth}{%
      \vspace{1mm}
      \textbf{ERQ1:}
        Which specific code smell types in a class influence its stability?
      \vspace{1mm}
    }%
    \hspace{3pt}%
  }%
}

\vspace{3mm}

\noindent
$\boldsymbol{H1.1_{[\mathrm{CS}]}}$: Classes exhibiting the \textbf{code smell type $\mathrm{CS}$} are significantly \textbf{less stable} than classes not exhibiting that smell type.

\noindent
$\boldsymbol{H1.2_{[\mathrm{CS}]}}$: There is a significant \textbf{negative association} between the \textbf{number of instances of the code smell type $\mathrm{CS}$} a class exhibits and its stability.

For $\boldsymbol{H1.2_{[\mathrm{CS}]}}$, we examine only method-level smells because the count of a class-level smell type is effectively binary at the focal-class level and would duplicate the presence analysis in $\boldsymbol{H1.1_{[\mathrm{CS}]}}$.

\subsubsection{RQ2}

As \citet{martin1994oo, martin2000design} suggests, instability is proportional to efferent coupling. Our theory is that if the efferent neighbors exhibit code smells, the focal class becomes less stable than if the efferent neighbors are clean. This is because prior empirical studies suggest that smelly classes tend to change more frequently \citep{santana2024unraveling, khomh2009exploratory, palomba2018diffuseness, olbrich2009evolution}, which in turn may lead the classes depending on them to change more often. Such a situation might even be true for non-smelly classes. Therefore, we test both the population of all classes and the population of non-smelly classes to isolate the effect of smells in efferent neighbors from the effect of smells in the focal class. As in RQ1, we also consider the presence, number, and variety of code smells, but here in efferent neighbors rather than in the focal class. Hence, our RQ2 and its hypotheses are:

\vspace{1mm}
\noindent
\fbox{%
  \parbox{\columnwidth}{%
    \hspace{3pt}%
    \parbox{0.95\columnwidth}{%
      \vspace{1mm}
      \textbf{RQ2:}
        Do code smells in a class's efferent neighbors influence its stability?
      \vspace{1mm}
    }%
    \hspace{3pt}%
  }%
}

\vspace{3mm}

\noindent
$\boldsymbol{H2.1}$: Classes with \textbf{smelly efferent neighbors} are significantly \textbf{less stable} than classes without.

\noindent
$\boldsymbol{H2.2}$: There is a significant \textbf{negative association} between the \textbf{number of code smells} in a class's efferent neighbors and its stability.

\noindent
$\boldsymbol{H2.3}$: There is a significant \textbf{negative association} between the \textbf{variety of code smells} in a class's efferent neighbors and its stability.

\medskip

\noindent
$\boldsymbol{H2.4}$: \underline{\emph{Non-smelly}} classes with \textbf{smelly efferent neighbors} are significantly \textbf{less stable} than those without.

\noindent
$\boldsymbol{H2.5}$: There is a significant \textbf{negative association} between the \textbf{number of code smells} in a \underline{\emph{non-smelly}} class's efferent neighbors and its stability.

\noindent
$\boldsymbol{H2.6}$: There is a significant \textbf{negative association} between the \textbf{variety of code smells} in a \underline{\emph{non-smelly}} class's efferent neighbors and its stability.

\medskip

RQ2 also aggregates the measures across all smell types. To explore which specific code smell types in efferent neighbors are associated with stability, we further examine $\boldsymbol{H2.1}$, $\boldsymbol{H2.2}$, $\boldsymbol{H2.4}$, and $\boldsymbol{H2.5}$ for each code smell type. We do not consider $\boldsymbol{H2.3}$ or $\boldsymbol{H2.6}$ because the variety of a single selected smell type reduces to whether that smell type is present, which is already addressed by $\boldsymbol{H2.1}$ and $\boldsymbol{H2.4}$, respectively. Thus, our ERQ2 and its exploratory hypotheses are:

\vspace{1mm}
\noindent
\fbox{%
  \parbox{\columnwidth}{%
    \hspace{3pt}%
    \parbox{0.95\columnwidth}{%
      \vspace{1mm}
      \textbf{ERQ2:}
        Which specific code smell types in a class's efferent neighbors influence its stability?
      \vspace{1mm}
    }%
    \hspace{3pt}%
  }%
}

\vspace{3mm}

\noindent
$\boldsymbol{H2.1_{[\mathrm{CS}]}}$: Classes with efferent neighbors exhibiting the \textbf{code smell type $\mathrm{CS}$} are significantly \textbf{less stable} than classes without.

\noindent
$\boldsymbol{H2.2_{[\mathrm{CS}]}}$: There is a significant \textbf{negative association} between the \textbf{number of instances of the code smell type $\mathrm{CS}$} in a class's efferent neighbors and its stability.

\medskip

\noindent
$\boldsymbol{H2.4_{[\mathrm{CS}]}}$: \underline{\emph{Non-smelly}} classes with efferent neighbors exhibiting the \textbf{code smell type $\mathrm{CS}$} are significantly \textbf{less stable} than those without.

\noindent
$\boldsymbol{H2.5_{[\mathrm{CS}]}}$: There is a significant \textbf{negative association} between the \textbf{number of instances of the code smell type $\mathrm{CS}$} in a \underline{\emph{non-smelly}} class's efferent neighbors and its stability.

\subsubsection{RQ3} \label{subsubsec:rq3_details}

We speculate that efferent code smell coupling may be worse for stability than cases where (a) only the focal class is smelly, (b) only the efferent neighbors are smelly, or (c) neither is smelly. Our theory is that when a code smell instance in an efferent neighbor changes, the ripple effect through the dependency may also require developers to change the code smell instance in the focal class. The potential need to change code smell instances in both the focal class and an efferent neighbor \emph{at once} may lead to more frequent and larger changes to the focal class, thereby further reducing its stability. In particular, we examine whether the joint presence of code smells in the focal class and its efferent neighbors is associated with class stability beyond their separate associations. We therefore test this association both without accounting for the separate presence of code smells in the two locations and after accounting for them. Hence, our RQ3 and its hypotheses are:

\vspace{1mm}
\noindent
\fbox{%
  \parbox{\columnwidth}{%
    \hspace{3pt}%
    \parbox{0.95\columnwidth}{%
      \vspace{1mm}
      \textbf{RQ3:}
        Does efferent code smell coupling influence a class's stability?
      \vspace{1mm}
    }%
    \hspace{3pt}%
  }%
}

\vspace{3mm}

\noindent
$\boldsymbol{H3.1}$: Classes with \textbf{efferent code smell coupling} are significantly \textbf{less stable} than classes without, if the presence of code smells in the class itself and in its efferent neighbors are \underline{ignored}.

\medskip

\noindent
$\boldsymbol{H3.2}$: Classes with \textbf{efferent code smell coupling} are significantly \textbf{less stable} than classes without, if the presence of code smells in the class itself and in its efferent neighbors are \underline{held constant}.

\subsubsection{RQ4} \label{subsubsec:rq4_details}

We speculate that efferent code smell interaction may further influence stability, as the code smell instances in the focal class and an efferent neighbor are directly connected by a static dependency. In such cases, when the smell in the efferent neighbor changes, the dependency makes it more likely that developers must also change the corresponding smell instance in the focal class. As a result, developers are more likely to need to modify smell instances in two classes at once. In contrast, in efferent code smell coupling without a direct static dependency between the smell instances, the probability of needing to change smells in both classes at once is lower. Additionally, since interacting code smells are associated with more static dependencies between them \citep{zhang2025analyzing}, there are more routes for changes to ripple into the focal class, making it more likely to be changed more often and more extensively. Prior work also shows that a higher number of such dependencies can harm comprehensibility and affect a developer's ability to effectively perform modifications \citep{yamashita2013exploring}. Hence, our RQ4 and its hypotheses are:

\vspace{1mm}
\noindent
\fbox{%
  \parbox{\columnwidth}{%
    \hspace{3pt}%
    \parbox{0.95\columnwidth}{%
      \vspace{1mm}
      \textbf{RQ4:}
        Does efferent code smell interaction influence a class's stability?
      \vspace{1mm}
    }%
    \hspace{3pt}%
  }%
}

\vspace{3mm}

\noindent
$\boldsymbol{H4.1}$: Classes with \textbf{efferent code smell interactions} are significantly \textbf{less stable} than classes without, when the number of code smells they exhibit and the number of code smells in their efferent neighbors are \underline{ignored}.

\noindent
$\boldsymbol{H4.2}$: There is a significant \textbf{negative association} between the \textbf{number of efferent code smell interactions} in a class and its stability, when the number of code smells it exhibits and those in its efferent neighbors are \underline{ignored}.

\noindent
$\boldsymbol{H4.3}$: There is a significant \textbf{negative association} between the \textbf{intensity of efferent code smell interactions} in a class and its stability, when the number of code smells it exhibits and those in its efferent neighbors are \underline{ignored}.

\medskip

\noindent
$\boldsymbol{H4.4}$: Classes with \textbf{efferent code smell interactions} are significantly \textbf{less stable} than classes without, when the number of code smells they exhibit and the number of code smells in their efferent neighbors are \underline{held constant}.

\noindent
$\boldsymbol{H4.5}$: There is a significant \textbf{negative association} between the \textbf{number of efferent code smell interactions} in a class and its stability, when the number of code smells it exhibits and those in its efferent neighbors are \underline{held constant}.

\noindent
$\boldsymbol{H4.6}$: There is a significant \textbf{negative association} between the \textbf{intensity of efferent code smell interactions} in a class and its stability, when the number of code smells it exhibits and those in its efferent neighbors are \underline{held constant}.

\medskip

RQ4 aggregates the measures across all smell-type pairs. To explore the influence of each smell-type pair, we further examine $\boldsymbol{H4.4}$ to $\boldsymbol{H4.6}$ for each smell-type pair. The numbers of code smells in the focal class and its efferent neighbors are held constant to reduce confounding. Thus, our ERQ4 and its exploratory hypotheses are:

\vspace{1mm}
\noindent
\fbox{%
  \parbox{\columnwidth}{%
    \hspace{3pt}%
    \parbox{0.95\columnwidth}{%
      \vspace{1mm}
      \textbf{ERQ4:}
        Which smell-type pairs influence a class's stability through efferent code smell interaction?
      \vspace{1mm}
    }%
    \hspace{3pt}%
  }%
}

\vspace{3mm}

\noindent
$\boldsymbol{H4.4_{[\mathrm{CS}_1\rightarrow\mathrm{CS}_2]}}$: Classes with \textbf{efferent code smell interactions for the smell-type pair $\mathrm{CS}_1\rightarrow\mathrm{CS}_2$} are significantly \textbf{less stable} than classes without, when the number of code smells they exhibit and the number of code smells in their efferent neighbors are \underline{held constant}.

\noindent
$\boldsymbol{H4.5_{[\mathrm{CS}_1\rightarrow\mathrm{CS}_2]}}$: There is a significant \textbf{negative association} between the \textbf{number of efferent code smell interactions for the smell-type pair $\mathrm{CS}_1\rightarrow\mathrm{CS}_2$} in a class and its stability, when the number of code smells it exhibits and those in its efferent neighbors are \underline{held constant}.

\noindent
$\boldsymbol{H4.6_{[\mathrm{CS}_1\rightarrow\mathrm{CS}_2]}}$: There is a significant \textbf{negative association} between the \textbf{intensity of efferent code smell interactions for the smell-type pair $\mathrm{CS}_1\rightarrow\mathrm{CS}_2$} in a class and its stability, when the number of code smells it exhibits and those in its efferent neighbors are \underline{held constant}.

\medskip

For the exploratory analyses, $\mathrm{CS}$ ranges over the 10 code smell types considered in this study, except for $\boldsymbol{H1.2_{[\mathrm{CS}]}}$, where it ranges over the 5 method-level smell types. $\mathrm{CS}_1$ and $\mathrm{CS}_2$ each range over the 10 code smell types, yielding 100 smell-type pairs $\mathrm{CS}_1\rightarrow\mathrm{CS}_2$.

For readability, when discussing the exploratory results, we omit the smell-type or smell-type-pair subscript from the hypothesis notation when the ERQ context is clear. For example, $\boldsymbol{H1.1_{[\mathrm{CS}]}}$ is referred to as $\boldsymbol{H1.1}$ in ERQ1, and $\boldsymbol{H4.4_{[\mathrm{CS}_1\rightarrow\mathrm{CS}_2]}}$ is referred to as $\boldsymbol{H4.4}$ in ERQ4.

\subsection{Independent Variables (IVs)} \label{sec:indep_variable}

We define the following independent variables (\emph{\textbf{IVs}}):

\begin{itemize}[leftmargin=0pt, itemsep=0.1em]
    \item \textbf{Is the focal class smelly (\texttt{IsSmelly})}: Indicates whether the focal class exhibits at least one code smell. $\texttt{IsSmelly}_{[\mathrm{CS}]}$ indicates whether the focal class exhibits the code smell type $\mathrm{CS}$.
    
    \item \textbf{Number of code smells in the focal class (\texttt{\#SmellFoc})}: The number of code smell instances in the focal class. $\texttt{\#SmellFoc}_{[\mathrm{CS}]}$ denotes the number of instances of the code smell type $\mathrm{CS}$ in the focal class.

    \item 
    \textbf{Variety of code smells in the focal class (\texttt{VarSmellFoc})}: The number of different types of code smells present in the focal class.
    
    \item \textbf{Has smelly efferent neighbor (\texttt{HasSmellEff})}: Indicates whether at least one efferent neighbor of the focal class exhibits a code smell. $\texttt{HasSmellEff}_{[\mathrm{CS}]}$ indicates whether at least one efferent neighbor exhibits the code smell type $\mathrm{CS}$.
    
    \item \textbf{Number of smells in efferent neighbors (\texttt{\#SmellEff})}: The total number of code smell instances across all efferent neighbors of the focal class. $\texttt{\#SmellEff}_{[\mathrm{CS}]}$ denotes the number of instances of the code smell type $\mathrm{CS}$ across these efferent neighbors.

    \item 
    \textbf{Variety of smells in efferent neighbors (\texttt{VarSmellEff})}: The number of different types of code smells present across all efferent neighbors of the focal class.
    
    \item \textbf{Has efferent code smell coupling (\texttt{HasEffCoup})}: Indicates whether the focal class and at least one of its efferent neighbors exhibit code smells. This is equivalent to \texttt{HasEffCoup} $=$ \texttt{IsSmelly} $\times$ \texttt{HasSmellEff}.
    
    \item \textbf{Has efferent code smell interaction (\texttt{HasEffInt})}: Indicates whether the focal class exhibits an efferent code smell interaction. $\texttt{HasEffInt}_{[\mathrm{CS}_1\rightarrow\mathrm{CS}_2]}$ indicates whether an efferent code smell interaction occurs for the smell-type pair $\mathrm{CS}_1\rightarrow\mathrm{CS}_2$.
    
    \item \textbf{Number of efferent code smell interactions (\texttt{\#EffInt})}: The number of distinct EffInt smell-instance pairs for the focal class. $\texttt{\#EffInt}_{[\mathrm{CS}_1\rightarrow\mathrm{CS}_2]}$ denotes the number of such instance pairs belonging to the smell-type pair $\mathrm{CS}_1\rightarrow\mathrm{CS}_2$.
    
    \item \textbf{Efferent code smell interaction intensity (\texttt{EffIntInten})}: The total number of EffInt dependency instances connecting EffInt smell-instance pairs. $\texttt{EffIntInten}_{[\mathrm{CS}_1\rightarrow\mathrm{CS}_2]}$ denotes the corresponding interaction intensity for the smell-type pair $\mathrm{CS}_1\rightarrow\mathrm{CS}_2$.
\end{itemize}

Since our unit of analysis is the class, we attribute both class-level and method-level smell instances to their enclosing class (i.e., each method-level smell instance is assigned to the class that contains the method). We do not apply weights based on smell level or smell type. In our exploratory analyses, we separately examine individual code smell types and directional smell-type pairs to explore their associations with stability.

\subsection{Dependent Variables (DVs)}
\label{sec:dep_variable}

In this section, we introduce the metrics used to measure class stability, which serve as our dependent variables (\emph{\textbf{DVs}}). For each focal class, we measure its stability over the observation period using the following metrics:

\begin{itemize}[leftmargin=0pt, itemsep=0.1em]
    \item \textbf{Change Frequency (\texttt{ChF})}: The number of commits that modify the class during the period. Commits involving only changes to comments, blank lines, whitespace, or indentation are not counted.

    \item \textbf{Change Size (\texttt{ChS})}: The total number of lines of code added or deleted in the class, excluding blank lines, comments, and changes involving only whitespace or indentation, aggregated over all commits during the period.
\end{itemize}

These two metrics capture complementary aspects of stability: \texttt{ChF} reflects how often a class changes, while \texttt{ChS} reflects how much it changes. Using both provides a more complete view of class stability. 

Since stability is measured with two metrics, each hypothesis listed in Section \ref{sec:rq_hypo} is treated as two separate hypotheses. We add an additional superscript to indicate which metric the hypothesis is testing. For example, $\boldsymbol{H1.1}$ is treated as two hypotheses: $\boldsymbol{H1.1^{\text{\texttt{ChF}}}}$ and $\boldsymbol{H1.1^{\text{\texttt{ChS}}}}$.

\subsection{Control Variables (CVs)}

For all hypotheses, we include the following control variables (\emph{\textbf{CVs}}):

\begin{itemize}[leftmargin=0pt, itemsep=0.1em]
    \item \textbf{Class size (\texttt{ClSize})}: The lines of code in the class, excluding comments and blank lines.

    \item \textbf{Number of efferent neighbors (\texttt{\#EffNei})}: The number of classes on which the focal class depends (i.e., its efferent coupling).
\end{itemize}

The rationale for including these controls is to reduce confounding. Larger classes inherently have a higher probability of being changed \citep{olbrich2010all}, and classes with more efferent neighbors are also more likely to change because modifications to any of their efferent neighbors may trigger changes in the focal class \citep{martin1994oo, martin2000design}.

For hypothesis $\boldsymbol{H3.2}$, we additionally control for \texttt{IsSmelly} and \texttt{HasSmellEff}. For hypotheses $\boldsymbol{H4.4}$ to $\boldsymbol{H4.6}$ and $\boldsymbol{H4.4_{[\mathrm{CS}_1\rightarrow\mathrm{CS}_2]}}$ to $\boldsymbol{H4.6_{[\mathrm{CS}_1\rightarrow\mathrm{CS}_2]}}$, we additionally control for \texttt{\#SmellFoc} and \texttt{\#SmellEff}. The rationale is explained in Section~\ref{sec:rq_hypo}. We do not include \texttt{VarSmellFoc} and \texttt{VarSmellEff} as control variables in these hypotheses. These variables capture smell-type diversity and are likely to be correlated with \texttt{\#SmellFoc} and \texttt{\#SmellEff}. Including both count and variety may introduce multicollinearity and reduce interpretability. Because the primary purpose of the additional controls is to adjust for baseline smell conditions, we treat \texttt{\#SmellFoc} and \texttt{\#SmellEff} as reasonable proxies for the main deconfounding goal.

When focusing on one specific smell type or smell-type pair in the exploratory analyses, we include one additional control representing the corresponding remaining context outside the selected smell type or pair, to better isolate its association with class stability. Specifically, when analyzing a code smell type, we account for the remaining smell context in the focal class or its efferent neighbors; when analyzing a smell-type pair, we account for the corresponding remaining efferent code smell interaction context. We define these controls as follows:

\begin{itemize}[leftmargin=0pt, itemsep=0.1em]
    \item \textbf{Has other code smells in the focal class ($\texttt{HasOtherSmellFoc}_{[\mathrm{CS}]}$)}: Indicates whether the focal class contains any code smell instances other than those of the selected smell type $\mathrm{CS}$.

    \item \textbf{Number of other code smells in the focal class ($\texttt{\#OtherSmellFoc}_{[\mathrm{CS}]}$)}: The number of code smell instances in the focal class other than those of the selected smell type $\mathrm{CS}$, defined as $\texttt{\#OtherSmellFoc}_{[\mathrm{CS}]}=\texttt{\#SmellFoc}-\texttt{\#SmellFoc}_{[\mathrm{CS}]}$.

    \item \textbf{Has other code smells in efferent neighbors ($\texttt{HasOtherSmellEff}_{[\mathrm{CS}]}$)}: Indicates whether the focal class's efferent neighbors contain any code smell instances other than those of the selected smell type $\mathrm{CS}$.

    \item \textbf{Number of other code smells in efferent neighbors ($\texttt{\#OtherSmellEff}_{[\mathrm{CS}]}$)}: The number of code smell instances in the focal class's efferent neighbors other than those of the selected smell type $\mathrm{CS}$, defined as $\texttt{\#OtherSmellEff}_{[\mathrm{CS}]}=\texttt{\#SmellEff}-\texttt{\#SmellEff}_{[\mathrm{CS}]}$.

    \item \textbf{Has other efferent code smell interaction ($\texttt{HasOtherEffInt}_{[\mathrm{CS}_1\rightarrow\mathrm{CS}_2]}$)}: Indicates whether the class exhibits any EffInt smell-instance pairs outside the selected smell-type pair $\mathrm{CS}_1\rightarrow\mathrm{CS}_2$.

    \item \textbf{Number of other efferent code smell interactions ($\texttt{\#OtherEffInt}_{[\mathrm{CS}_1\rightarrow\mathrm{CS}_2]}$)}: The number of distinct EffInt smell-instance pairs outside the selected smell-type pair $\mathrm{CS}_1\rightarrow\mathrm{CS}_2$, defined as $\texttt{\#OtherEffInt}_{[\mathrm{CS}_1\rightarrow\mathrm{CS}_2]}=\texttt{\#EffInt}-\texttt{\#EffInt}_{[\mathrm{CS}_1\rightarrow\mathrm{CS}_2]}$.

    \item \textbf{Other efferent code smell interaction intensity ($\texttt{OtherEffIntInten}_{[\mathrm{CS}_1\rightarrow\mathrm{CS}_2]}$)}: The EffInt intensity outside the selected smell-type pair, defined as $\texttt{OtherEffIntInten}_{[\mathrm{CS}_1\rightarrow\mathrm{CS}_2]} = \texttt{EffIntInten}-\texttt{EffIntInten}_{[\mathrm{CS}_1\rightarrow\mathrm{CS}_2]}$.
\end{itemize}

Table \ref{tab:hypo_variable} lists the IVs and CVs for each hypothesis. The DVs are always the two listed in Section \ref{sec:dep_variable}.

\begin{table}[t]
    \centering
    \caption{Independent and control variables per hypothesis}
    \renewcommand{\arraystretch}{1.1}
    \begin{tabular}{c|c|c}
        \hline
        \textbf{Hypotheses} & \textbf{Independent Variables} & \textbf{Control Variables} \\ 
        \hline
        $\boldsymbol{H1.1}$ & IsSmelly & ClSize, \#EffNei \\
        \hline
        $\boldsymbol{H1.1}_{[\mathrm{CS}]}$ & $\texttt{IsSmelly}_{[\mathrm{CS}]}$ & \makecell{ClSize, \#EffNei, \\ $\texttt{HasOtherSmellFoc}_{[\mathrm{CS}]}$} \\

        \hline
        $\boldsymbol{H1.2}$ & \#SmellFoc & ClSize, \#EffNei \\
        \hline
        $\boldsymbol{H1.2}_{[\mathrm{CS}]}$ & $\texttt{\#SmellFoc}_{[\mathrm{CS}]}$ & \makecell{ClSize, \#EffNei, \\ $\texttt{\#OtherSmellFoc}_{[\mathrm{CS}]}$} \\

        \hline
        $\boldsymbol{H1.3}$ & VarSmellFoc & ClSize, \#EffNei \\

        \hline
        $\boldsymbol{H2.1}$ \& $\boldsymbol{H2.4}$ & HasSmellEff & ClSize, \#EffNei \\ 
        \hline
        $\boldsymbol{H2.1}_{[\mathrm{CS}]}$ \& $\boldsymbol{H2.4}_{[\mathrm{CS}]}$ & $\texttt{HasSmellEff}_{[\mathrm{CS}]}$ & \makecell{ClSize, \#EffNei, \\ $\texttt{HasOtherSmellEff}_{[\mathrm{CS}]}$} \\ 

        \hline
        $\boldsymbol{H2.2}$ \& $\boldsymbol{H2.5}$ & \#SmellEff & ClSize, \#EffNei \\
        \hline
        $\boldsymbol{H2.2}_{[\mathrm{CS}]}$ \& $\boldsymbol{H2.5}_{[\mathrm{CS}]}$ & $\texttt{\#SmellEff}_{[\mathrm{CS}]}$ & \makecell{ClSize, \#EffNei, \\ $\texttt{\#OtherSmellEff}_{[\mathrm{CS}]}$} \\

        \hline
        $\boldsymbol{H2.3}$ \& $\boldsymbol{H2.6}$ & VarSmellEff & ClSize, \#EffNei \\

        \hline
        $\boldsymbol{H3.1}$ & HasEffCoup & ClSize, \#EffNei \\ 

        \hline
        $\boldsymbol{H3.2}$ & HasEffCoup & \makecell{ClSize, \#EffNei, \\ IsSmelly, HasSmellEff} \\ 

        \hline
        $\boldsymbol{H4.1}$ & HasEffInt & ClSize, \#EffNei \\

        \hline
        $\boldsymbol{H4.2}$ & \#EffInt & ClSize, \#EffNei \\

        \hline
        $\boldsymbol{H4.3}$ & EffIntInten & ClSize, \#EffNei \\ 

        \hline
        $\boldsymbol{H4.4}$ & HasEffInt & \makecell{ClSize, \#EffNei, \\ \#SmellFoc, \#SmellEff} \\
        \hline
        $\boldsymbol{H4.4}_{[\mathrm{CS}_1\rightarrow\mathrm{CS}_2]}$ & $\texttt{HasEffInt}_{[\mathrm{CS}_1\rightarrow\mathrm{CS}_2]}$ & \makecell{ClSize, \#EffNei, \\ \#SmellFoc, \#SmellEff, \\ $\texttt{HasOtherEffInt}_{[\mathrm{CS}_1\rightarrow\mathrm{CS}_2]}$} \\
        \hline

        $\boldsymbol{H4.5}$ & \#EffInt & \makecell{ClSize, \#EffNei, \\ \#SmellFoc, \#SmellEff} \\
        \hline
        $\boldsymbol{H4.5}_{[\mathrm{CS}_1\rightarrow\mathrm{CS}_2]}$ & $\texttt{\#EffInt}_{[\mathrm{CS}_1\rightarrow\mathrm{CS}_2]}$ & \makecell{ClSize, \#EffNei, \\ \#SmellFoc, \#SmellEff, \\ $\texttt{\#OtherEffInt}_{[\mathrm{CS}_1\rightarrow\mathrm{CS}_2]}$} \\
        \hline

        $\boldsymbol{H4.6}$ & EffIntInten & \makecell{ClSize, \#EffNei, \\ \#SmellFoc, \#SmellEff} \\
        \hline
        $\boldsymbol{H4.6}_{[\mathrm{CS}_1\rightarrow\mathrm{CS}_2]}$ & $\texttt{EffIntInten}_{[\mathrm{CS}_1\rightarrow\mathrm{CS}_2]}$ & \makecell{ClSize, \#EffNei, \\ \#SmellFoc, \#SmellEff, \\ $\texttt{OtherEffIntInten}_{[\mathrm{CS}_1\rightarrow\mathrm{CS}_2]}$} \\
        \hline
    \end{tabular}
    \label{tab:hypo_variable}
\end{table}

\subsection{Repository Mining} \label{sec:repo_mining}

For each system under consideration, we first selected a snapshot that had at least one year of subsequent commit history, which served as the observation window. Within each repository, we excluded third-party libraries because their source code is generally not maintained by the project developers, and insights derived from such code may therefore be less actionable for them. Additionally, we excluded test and example code to focus the analysis on production source code. For each class in that version, we obtained its size and the number of efferent neighbors, which served as the CVs (\texttt{ClSize} and \texttt{\#EffNei}). We then detected code smells in the class and its efferent neighbors, and obtained measurements of code smells, code smell coupling, and code smell interaction. These constituted the IVs and also served as additional CVs for certain hypotheses. Next, for each commit (including direct commits and pull-request merges) in the one-year period on the main branch, we identified all classes modified by the commit and computed the amount of code change for each class. Specifically, for each commit, we obtained the \emph{diff} for all modified classes and counted the added and deleted lines of code after removing blank lines, comments, and changes involving only whitespace or indentation. We aggregated these values across all commits to compute \texttt{ChS} and counted the commits with a positive amount of such code change to compute \texttt{ChF}.

We assume a one-to-one mapping while mining the commit history: (i) the focal class in the snapshot can be matched to exactly one class throughout the mining period, and (ii) each commit's changes can be attributed to that same single focal class in the snapshot over time. This assumption ensures that the IV and CVs measured for a focal class at the snapshot can be unambiguously associated with its DV computed from aggregated changes, and that each DV value can be unambiguously associated with the IV and CVs of a single focal class in the snapshot.

Classes that were renamed or moved did not violate this assumption; they were treated as the same evolving class and were retained in the analysis. Classes that were deleted during the observation period were also retained because the deletion is itself a change to the class and can still be unambiguously attributed to the corresponding focal class in the snapshot. Excluding deleted classes would systematically remove deletion events from the stability measure, even though such events can represent a substantial form of instability. For a deleted class, the commit that removes it contributes to ChF, while the lines removed by the deletion can still contribute to ChS. We excluded classes involved in class merge or class split refactorings during the observation period because these operations violate the one-to-one mapping, either by making the focal class no longer unambiguously attributable to the entity whose changes are aggregated (i.e., class split), or by making the entity whose changes are aggregated no longer unambiguously attributable to a single focal class in the snapshot (i.e., class merge). We did not apply special handling to other refactoring commits; such commits do not violate our assumption and were intentionally included because our stability measures are designed to capture all changes, regardless of whether they arise from feature work, bug fixes, or refactoring.

The rationale for choosing a one-year time span is to avoid large changes in the function of the class between the time of code smell detection and the commit-mining period, since such changes may alter the class's intrinsic stability.

It is possible that new smells are introduced or that smells are refactored from the focal class or its efferent neighbors during the period, which may influence code smell coupling and interaction. Our goal is to investigate class instability influenced by smells, smell couplings, and smell interactions, rather than explaining why smells were introduced or removed. Therefore, we do not assume that smells, smell coupling, or smell interaction remain constant during the period.

\subsection{System Selection}
\label{sec:system_selection}

We conducted our study on top-starred open-source projects from GitHub. The rationale for using top-starred projects is to prioritize the most popular and widely adopted projects \citep{santana2024unraveling}. We therefore ranked GitHub projects in descending order by number of stars. Following recommendations from \citet{kalliamvakou2016depth}, we applied the following inclusion criteria (IC): \textbf{IC1}. More than 100 forks (indicates meaningful developer interest). \textbf{IC2}. At least 20 contributors (to avoid personal or low-activity projects). \textbf{IC3}. At least one year of commit history with at least 50 commits within that one-year period (to ensure meaningful development activity, approximately one commit per week). \textbf{IC4}. More than 80\% of the code written in Java (due to the smell detection tool limitation). \textbf{IC5}. Not an educational project (educational projects may not represent real software development processes).

We selected the first 300 projects that satisfied IC1, IC2, and IC4. After cloning these projects, we analyzed each project's local commit history to apply IC3. This removed 145 projects, leaving 155 projects. We then manually applied IC5, which removed three additional projects. This process resulted in a final dataset of 152 projects.

Table~\ref{tab:proj_summary_stat} provides a statistical summary of the selected Java projects in terms of baseline project size and maintenance activity during the one-year observation period. Project size is characterized using lines of code (LOC) in Java files, excluding comments and blank lines, while maintenance activity during the observation window is characterized using the number of commits (\#Commits) and system-level churn (SysChurn), with SysChurn defined as the total number of lines added and deleted across all repository files and commits.

\begin{table}[tb]
    \centering
    \caption{Project size and maintenance activity summary statistics}
    \renewcommand{\arraystretch}{1.25}
    \begin{tabular}{lllllll}
        \hline
        & Mean & Median & Max & Min & Sum & SD \\
        \hline
        LOC & 201.9K & 98.7K & 1.5M & 7.5K & 30.7M & 266.9K \\
        \#Commits & 773.4 & 346.5 & 12.7K & 53 & 117.6K & 1.3K \\
        SysChurn & 372.4K & 118.5K & 5.3M & 2.8K & 56.6M & 806.6K \\
        \hline
        \multicolumn{7}{l}{K = thousand; M = million; SD = standard deviation} \\
    \end{tabular}
    \label{tab:proj_summary_stat}
\end{table}

The projects vary substantially in size, ranging from 7.5K LOC (Retrofit) to 1.5M LOC (Elasticsearch). Maintenance activity also varies across projects, with the number of commits ranging from 53 (Web3j) to 12.7K (Elasticsearch) and system-level churn ranging from 2.8K lines (MyBatis) to 5.3M lines (Bazel). The relatively large standard deviations for these measures further indicate substantial between-project variation in project size and maintenance activity. Overall, the selected systems span a wide range of project sizes and maintenance activity, providing a diverse empirical context for the subsequent analyses.

\subsection{Data Analysis}
\label{sec:data_analysis}

In this section, we describe how we analyzed the data. The analysis methods and statistical models discussed in this section apply to all our hypotheses.

Both DVs (\texttt{ChF} and \texttt{ChS}) are discrete count data and therefore take non-negative integer values. These characteristics make ordinary linear regression less suitable for our outcomes. We therefore used generalized linear mixed models (GLMMs), which can accommodate different forms of outcome data by specifying an appropriate distribution and a link function relating the predictors to the expected outcome \citep{gelman2007data}. GLMMs have also been used to analyze software engineering data \citep{kuutila2021individual,lambiase2024empirical}. For each hypothesis, we fitted a separate GLMM with the corresponding DV (either \texttt{ChF} or \texttt{ChS}) as the outcome and the IV and CVs specified for that hypothesis in Table~\ref{tab:hypo_variable} as predictors.

We modeled both DVs using negative binomial GLMMs with a log link. Poisson and negative binomial models are both standard approaches for count data, but the Poisson distribution assumes equidispersion, whereas software engineering count measures can exhibit overdispersion \citep{mccullagh2019generalized,ostrand2004bugs}. We additionally fitted corresponding Poisson GLMMs and computed dispersion statistics. The diagnostics indicated overdispersion for both \texttt{ChF} and \texttt{ChS}, supporting the use of negative binomial GLMMs.

We inspected histograms of the non-binary IVs and CVs and observed substantial right skew, with a small number of comparatively large values. To reduce the chance that a small number of unusually large observations disproportionately influence the fitted relationships \citep{changyong2014log}, we transformed the non-binary IVs and CVs using $\log(1+x)$, where $\log$ denotes the natural logarithm (base $e$) and $x$ is the original variable value. Adding 1 allows variables containing zero values to be transformed. Binary IVs and CVs were entered without transformation. The histograms of the DVs also showed right-skewed distributions, but we did not transform them because they are modeled directly using the negative binomial distribution with a log link.

GLMMs allow the model to include random effects. A random effect is a model component that represents systematic differences among groups, and random-effects models are increasingly used in empirical software engineering \citep{gelman2007data,kuutila2021individual,lambiase2024empirical}. In our study, we included a project-level random intercept in all models to account for systematic differences between projects, such as variation in age, development activity, commit practices, and team size. This allowed classes from the same project to share a project-specific baseline while estimating the associations between the predictors and class stability \citep{raudenbush2002hierarchical}. Using a random intercept rather than fixed effects is appropriate given the large number of projects and our aim to account for between-project variability while drawing inference beyond the specific projects in our dataset, rather than fitting a separate coefficient for each project \citep{gelman2007data}.

For each fitted model, we report the regression coefficient ($\beta$) for each IV together with its standard error, 95\% confidence interval, and $p$-value (computed using Wald tests for the regression coefficients; one-sided as specified by each directional hypothesis), which collectively describe the estimated direction of the association, its precision, and its statistical significance.

We report two effect size measures to convey the practical magnitude of the associations: Incidence Rate Ratios (IRRs) \citep{hilbe2011negative} and Average Marginal Effects (AMEs) \citep{long2006regression}. They provide complementary interpretations: IRRs express the associations multiplicatively, whereas AMEs express them additively. Table~\ref{tab:association_interpretation} summarizes the interpretation of the reported effect size measures for binary and non-binary IVs. For non-binary IVs, the IRR represents the multiplicative change in the expected outcome associated with a one-unit increase in $\log(1+x)$, where $x$ denotes the original IV value. This interpretation is not straightforward in terms of the original IV. We therefore additionally report $2^\beta$, which we refer to as the \emph{doubling association}, to provide a more intuitive multiplicative interpretation. The doubling association represents the multiplicative change in the expected outcome associated with doubling $1+\mathrm{IV}$. For readability, when interpreting this measure in the Results we refer to this as ``doubling the IV''; throughout the Results, this wording is shorthand for doubling $1+\mathrm{IV}$ as defined here.

\begin{table*}[t]
    \centering
    \caption{Interpretation of the effect size measures reported for binary and non-binary IVs.}
    \footnotesize
    \renewcommand{\arraystretch}{1.5}
    \begin{tabular}{@{}p{0.15\textwidth}@{\hspace{0.02\textwidth}}p{0.395\textwidth}@{\hspace{0.02\textwidth}}p{0.415\textwidth}@{}}
        \hline
        \textbf{Measure} & \textbf{Binary IV} & \textbf{Non-binary IV} \\
        \hline
        \textbf{IRR} &
        Multiplicative difference in the expected outcome when the IV changes from 0 to 1. For example, an IRR of 1.20 corresponds to a 20\% higher expected outcome for IV$=1$ than for IV$=0$. &
        Multiplicative change in the expected outcome associated with a one-unit increase in $\log(1+\mathrm{IV})$. For example, an IRR of 1.20 corresponds to a 20\% higher expected outcome for a one-unit increase in $\log(1+\mathrm{IV})$. \\
        \hline
        \textbf{AME} &
        Average additive change in the expected outcome when the IV changes from 0 to 1. For example, an AME of 2 indicates an average increase of two units in the expected outcome. &
        Average additive change in the expected outcome associated with a one-unit increase in the IV. For example, an AME of 2 indicates an average increase of two units in the expected outcome for a one-unit increase in the IV. \\
        \hline
        \textbf{Doubling association ($2^\beta$)} &
        N/A &
        Multiplicative change in the expected outcome when $1+\mathrm{IV}$ doubles. For example, a value of 1.20 corresponds to a 20\% higher expected outcome when $1+\mathrm{IV}$ doubles. \\
        \hline
    \end{tabular}
    \label{tab:association_interpretation}
\end{table*}

To assess model fit, we compared each fitted model with a corresponding null model containing only an overall intercept and the same project-level random intercept. We used the log-likelihood \citep{mccullagh2019generalized} to assess how well the models explain the observed data. In our models, log-likelihood values are negative, and a higher (i.e., less negative) value indicates better fit. We also computed McFadden's pseudo-$R^2$ \citep{mcfadden1972conditional} as $R^2_{\mathrm{McFadden}} = 1 - \frac{\log L_{\mathrm{fitted}}}{\log L_{\mathrm{null}}}$, where $\log L_{\mathrm{fitted}}$ and $\log L_{\mathrm{null}}$ are the log-likelihoods of the fitted and corresponding null models, respectively. A value of 0 indicates no improvement over the null model, while larger positive values indicate greater improvement in model fit. Together, these measures allow us to assess whether including the IV and CVs provides additional explanatory value beyond the null model.

We assessed multicollinearity among the fixed-effect predictors (i.e., the IVs and CVs) using variance inflation factors (VIFs). VIF values range from 1 upward, where a value of 1 indicates no linear association between a predictor and the remaining predictors, and larger values indicate greater multicollinearity. Following commonly used thresholds in regression analysis and empirical software engineering, we treated VIF values of $\leq 5$ as indicating an acceptable level of multicollinearity, values of $>5$ and $\leq 10$ as indicating elevated multicollinearity, and values of $>10$ as indicating severe multicollinearity \citep{hair2010multivariate,fox2008applied,jiarpakdee2018autospearman,laaber2021predicting}.

For the primary analyses, because each RQ is addressed via multiple hypotheses, evaluated for both stability outcomes (\texttt{ChF} and \texttt{ChS}), and each RQ corresponds to a distinct confirmatory claim about a different theoretical mechanism linking code smells to stability (e.g., mechanisms related to smells in the class, in efferent neighbors, efferent code smell coupling, and efferent code smell interaction), we defined the family of tests separately for each RQ. The study was not designed to evaluate a single claim that ``code smells are related to stability'' in general, but rather to test whether such a relationship is supported under specific mechanisms; accordingly, inferential statements were interpreted per RQ, and conclusions for one RQ were not used to support claims about another. We therefore controlled the false discovery rate using the Benjamini-Hochberg (BH) procedure separately within each RQ. Specifically, for each RQ, we adjusted together all one-sided $p$-values for the IV coefficients tested under that RQ. This resulted in adjustments over 6, 12, 4, and 12 $p$-values for RQ1, RQ2, RQ3, and RQ4, respectively.

For the exploratory analyses, we likewise defined a separate family of tests for each ERQ because each ERQ examines a distinct set of fine-grained hypotheses, and applied BH independently within each family. The BH correction included $10$ code smell types $\times~2$ stability outcomes $+\ 5$ method-level code smell types $\times~2=30$ $p$-values for ERQ1; $10\times4$ hypotheses $\times~2=80$ for ERQ2; and $10$ source code smell types $\times~10$ target code smell types $\times~3$ hypotheses $\times~2=600$ for ERQ4.

We evaluated all primary and exploratory hypotheses using one-sided tests, consistent with the directional predictions specified in the study design. Statistical significance was assessed at $\alpha = 0.05$: an alternative hypothesis was considered accepted if the coefficient ($\beta$) of the IV was in the predicted direction and its one-sided BH-adjusted $p$-value $< 0.05$; otherwise, we concluded there was insufficient evidence to accept it. Both unadjusted and BH-adjusted $p$-values are available in the replication package \citep{effint2026zhang}.

\subsection{Implementation}
\label{sec:implmentation_sec}

Projects satisfying IC1, IC2, and IC4 were identified, selected, and downloaded automatically using the GitHub API \citep{github_api}. IC3 was then applied using Git commands to analyze each project's commit history \citep{git_api}.

Code smells were detected using our own tool based on the static analyzer CodeQL \citep{codeql}. CodeQL was also used to detect and measure static dependencies, efferent neighbors, and class size. RefactoringMiner \citep{refactoring_miner} and JGit \citep{jgit_api} were used to analyze commit history, obtain diffs, and identify class changes and refactorings. Python was used to automate the overall process and process the outputs from these tools to obtain the IVs, CVs, and DVs for each included class. These variables were then passed from Python to R through the rpy2 package \citep{rpy2_api}, where we fitted the GLMMs via the glmmTMB package \citep{glmmtmb} and obtained the statistical outputs.

Our previous study implemented the detector for 5 code smells \citep{zhang2025analyzing}; in this study, we extended it to detect 5 additional smells, covering all 10 smell types considered here. The rationale for developing our own detector is discussed in Sections~\ref{sec:deviations} and~\ref{sec:threats}. The detector closely follows the popular and widely used metric-based detection strategies proposed by \citet{lanza2007object}, including the metric definitions and thresholds specified in those strategies, to improve transparency and detection accuracy. When a detection threshold was specified as a range, we selected the stricter bound to reduce false positives. Specifically, for metrics whose values are required to be smaller than a threshold, we selected the lower end of the specified range, while for metrics whose values are required to be larger than a threshold, we selected the upper end. We used these stricter bounds to focus on artifacts that more clearly satisfy the intended definitions and to reduce the risk of classifying an artifact as smelly when it does not meet those definitions. Details of the code smell and metric definitions, selected thresholds, and detection rules are available in our replication package \citep{effint2026zhang}.

\noindent\textbf{Evaluation.} We evaluated the validity of the detector's classifications for all 10 smell types. For each smell, we determined the sample size using the standard confidence interval for a proportion with 95\% confidence and a $\pm 5\%$ margin of error. We then used stratified proportional sampling by project: for each code smell type $CS$, the sample size $n_{i,CS}$ allocated to project $i$ was proportional to $N_{i,CS}$, where $N_{i,CS}$ is the number of detected smell instances of type $CS$ in project $i$; we then drew a simple random sample for each code smell type within each project \citep{cochran1977sampling}. This sampling design reduces variance and keeps each project's representation proportional to its contribution \citep{singh1996stratified}.

We manually reviewed the sampled smell instances. For each sampled instance, we first checked whether the metric measurements were correct and then checked whether the smell instance matched the corresponding definition described by \citet{lanza2007object}. The detector also reports source-code information behind each metric measurement. For example, for metrics used to detect Feature Envy (FE), it reports which foreign fields are accessed and the classes in which those fields are declared. This information allowed us to efficiently check the metric measurements directly against the source code. We manually checked 321, 366, 369, 378, 364, 361, 376, 376, 380, and 144 smell instances for BC, BM, DC, DP, FE, GC, IT, RB, SG, and TB, respectively. All sampled smell instances and their metric measurements were judged correct, and no false positives were observed in the sample. This indicates that the detected smells in the sample were consistent with their definitions.

We applied the same sampling and review procedure to artifacts classified as not having the corresponding smell. We manually checked 384, 385, 384, 385, 385, 384, 385, 383, 385, and 378 such artifacts for BC, BM, DC, DP, FE, GC, IT, RB, SG, and TB, respectively. All metric measurements were judged correct. Potential false negatives accounted for 4.43\%, 1.82\%, 5.99\%, 1.82\%, 1.56\%, 5.99\%, 0.52\%, 4.18\%, 1.56\%, and 2.38\% of the sampled artifacts, respectively.

\subsection{Deviations from the Stage 1 Registered Plan}
\label{sec:deviations}

During study execution, we made three deviations from our Stage 1 registered report \citep{zhang2026influence}. These deviations did not change the preregistered primary RQs, hypotheses, stability measures, or inferential decision criteria.

\noindent\textbf{Number of projects.} The registered plan specified selecting the first 100 projects that satisfied all inclusion criteria. In practice, IC3 could only be assessed after cloning a repository and examining its commit history, while IC5 required manual inspection. To construct the sample, we therefore first selected the top 300 candidates satisfying the criteria that could be applied before repository cloning and manual inspection (IC1, IC2, and IC4), and then applied IC3 and IC5. This process left 152 projects, all of which satisfied the same preregistered inclusion criteria and came from the same top-starred sampling frame. We retained all 152 eligible projects to use the full set of projects satisfying the registered inclusion criteria, rather than reducing the dataset after eligibility had been established. Retaining all 152 eligible projects also increased the amount of data available for analysis, reduced unnecessary loss of information, and improved the generalizability of the findings within our sampling frame. The inclusion criteria themselves were unchanged.

\noindent\textbf{Code smell detector.} The registered plan specified JSpIRIT \citep{jspirit}. During execution, we found that JSpIRIT was not suitable for a large-scale automated analysis. JSpIRIT is an Eclipse plug-in, and each project module must be analyzed manually and individually rather than processed in batch. This makes it difficult to apply to projects containing many modules. JSpIRIT also does not provide functionality to export its detection results to a file; detected smells are shown only in an Eclipse view. In addition, it does not report the fully qualified name of each detected code smell instance, a limitation also noted by \citet{santana2024unraveling}. Different instances with the same simple name can therefore create naming conflicts and may be incorrectly matched to artifacts from other analysis steps. We therefore built on the CodeQL-based detector developed in our previous study \citep{zhang2025analyzing}, which detects 5 smell types, and extended it to detect the additional 5 smell types required in this study. This allows us to process projects in batch and output the fully qualified name of each code smell instance, enabling each instance to be accurately identified and matched to artifacts from other analysis steps. This change affected the implementation of smell detection rather than the preregistered detection strategy: the resulting detector covers all 10 smell types and follows the same metric-based detection strategies of \citet{lanza2007object} specified in the registered plan. We manually evaluated the detector. Our replication package provides the detector source code and implementation details \citep{effint2026zhang}.

\noindent\textbf{Exploratory analyses.} We added ERQ1, ERQ2, and ERQ4 to obtain finer-grained insight into the aggregate relationships examined by the primary RQs. Because these analyses were not part of the Stage 1 registered plan, we explicitly treat them as exploratory rather than confirmatory. They do not replace or modify the preregistered primary analyses, and their multiple-testing corrections are applied separately from those of the primary RQs. Their findings are interpreted as exploratory evidence that refines the primary results.

\section{Results} \label{sec:results}

This section presents our results. We first describe the dataset characteristics and model diagnostics, and then present the results for each RQ and ERQ in subsections.

The final analysis population consists of 287,580 classes from 152 projects. Across these classes, we detected 114,742 code smell instances: BC (1,929), BM (7,540), DC (9,080), DP (22,058), FE (6,702), GC (5,932), IT (15,770), RB (16,694), SG (28,808), and TB (229). Table~\ref{tab:code_smells} provides the definitions of these code smell types.

We also examined model fit and multicollinearity using the diagnostics described in Section~\ref{sec:data_analysis}. For model fit, all fitted models have higher log-likelihoods than their corresponding null models and positive McFadden's pseudo-$R^2$ values, indicating improved fit over the null models.

For multicollinearity, the median VIF across all fixed-effect predictors (i.e., the IVs and CVs) in all models considered in this study is 2.73. Only two variables have VIF values $>5$, both in $\boldsymbol{H3.2}$: \texttt{HasEffCoup} (6.60) and \texttt{IsSmelly} (6.15). These values fall within the elevated multicollinearity range ($>5$ and $\leq10$), and no VIF value indicates severe multicollinearity ($>10$). The elevated collinearity in $\boldsymbol{H3.2}$ is expected because \texttt{HasEffCoup} represents the joint presence of code smells in the focal class (\texttt{IsSmelly}) and its efferent neighbors (\texttt{HasSmellEff}), and is defined as \texttt{HasEffCoup} $=$ \texttt{IsSmelly} $\times$ \texttt{HasSmellEff}. We therefore retain these predictors because their VIF values are only moderately above 5, this collinearity follows from the intended model specification, and removing them would prevent the model from distinguishing the joint association from the separate associations that $\boldsymbol{H3.2}$ is designed to test.

Due to the large number of fitted models, the complete log-likelihood, McFadden's pseudo-$R^2$, and VIF values for all models are available in the replication package \citep{effint2026zhang}.

\subsection{RQ1 and ERQ1: Code Smells in the Focal Class}
\label{sec:results_rq1_erq1}

\begin{figure*}[tb]
    \centering
    \includegraphics[width=\textwidth]{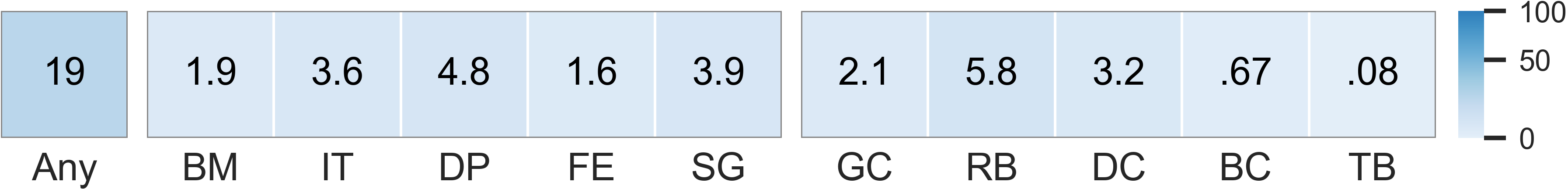}
    \caption{Prevalence of code smells in focal classes.}
    \label{fig:erq1_focal_prevalence}
\end{figure*}

We first present the prevalence of code smells in the focal class, measured as the percentage of analyzed classes that exhibit any code smell or a specific code smell type, as shown in the blue-hue heatmap in Figure~\ref{fig:erq1_focal_prevalence}. Of the 287,580 analyzed classes, 54,702 (19\%) exhibit at least one code smell. RB is the most prevalent smell (5.8\%), followed by DP (4.8\%), SG (3.9\%), IT (3.6\%), and DC (3.2\%). GC occurs in 2.1\% of classes, followed by BM (1.9\%), FE (1.6\%), BC (0.67\%), and TB (0.08\%). These percentages are not mutually exclusive because a class can exhibit multiple smell types.

\subsubsection{RQ1: Do code smells in a class influence its stability?}

Table~\ref{tab:rq1_results} presents the RQ1 results for the two measures of class stability, change frequency (\texttt{ChF}) and change size (\texttt{ChS}), as defined in Section~\ref{sec:dep_variable}. Here, $p$ denotes the BH-adjusted $p$-value, with statistically significant values ($p<0.05$) shown in bold. The complete results for all primary-analysis hypotheses, including $\beta$, its standard error, 95\% confidence interval (95\% CI), unadjusted $p$-value, and BH-adjusted $p$-value, are reported in Appendix~\ref{app:primary_results}, Table~\ref{tab:primary_results_full}.

\begin{table*}[tb]
    \centering
    \caption{RQ1 results.}
    \renewcommand{\arraystretch}{1.25}
    \small
    \begin{tabular}{lrrrr}
        \hline
        \textbf{Hypothesis} & \textbf{IRR} & \textbf{AME} & \makecell{\textbf{Doubling}\\\textbf{association}} & $\boldsymbol{p}$ \\
        \hline
        $\boldsymbol{H1.1^{\text{\texttt{ChF}}}}$ & 1.011 & 0.007 & -- & 0.191 \\
        $\boldsymbol{H1.1^{\text{\texttt{ChS}}}}$ & 0.982 & -0.213 & -- & 0.877 \\
        $\boldsymbol{H1.2^{\text{\texttt{ChF}}}}$ & 1.061 & 0.027 & 1.042 & $\boldsymbol{3.65\times10^{-15}}$ \\
        $\boldsymbol{H1.2^{\text{\texttt{ChS}}}}$ & 1.006 & 0.043 & 1.004 & 0.415 \\
        $\boldsymbol{H1.3^{\text{\texttt{ChF}}}}$ & 1.046 & 0.021 & 1.032 & $\boldsymbol{2.26\times10^{-6}}$ \\
        $\boldsymbol{H1.3^{\text{\texttt{ChS}}}}$ & 1.014 & 0.102 & 1.009 & 0.342 \\
        \hline
    \end{tabular}
    \label{tab:rq1_results}
\end{table*}

For $\boldsymbol{H1.1}$, the IRR of 1.011 corresponds to 1.1\% higher \texttt{ChF} for smelly than non-smelly classes, while the AME of 0.007 corresponds to an average increase of 0.007 in \texttt{ChF}. For $\boldsymbol{H1.2}$, the IRR of 1.061 corresponds to 6.1\% higher \texttt{ChF} for a one-unit increase in $\log(1+\texttt{\#SmellFoc})$, the AME of 0.027 corresponds to an average increase of 0.027 in \texttt{ChF} for one additional smell instance, and the doubling association of 1.042 corresponds to 4.2\% higher \texttt{ChF} when the number of smells doubles.

We find no evidence that simply being smelly is associated with lower stability: H1.1 is not supported for either \texttt{ChF} or \texttt{ChS}. In contrast, H1.2 and H1.3 are supported for \texttt{ChF}, but not for \texttt{ChS}. Doubling the number of smells is associated with 4.2\% higher \texttt{ChF} (doubling association = 1.042), while doubling the variety of smells is associated with 3.2\% higher \texttt{ChF} (doubling association = 1.032). Thus, at the aggregate level, the number and variety of smells are associated with change frequency, whereas none of the RQ1 measures shows a significant association with change size.

\subsubsection{ERQ1: Which specific code smell types in a class influence its stability?}

\begin{figure*}[tb]
    \centering
    \includegraphics[width=\textwidth]{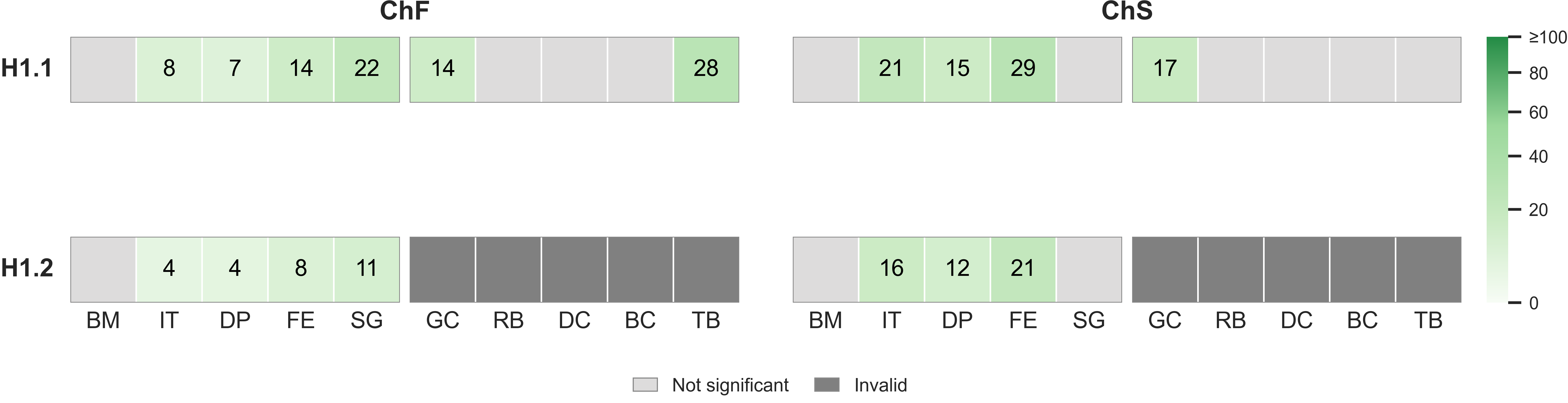}
    \caption{Percentage increases in \texttt{ChF} and \texttt{ChS} associated with code smells in focal classes.}
    \label{fig:erq1_effect_heatmap}
\end{figure*}

ERQ1 examines the corresponding smell-specific associations in the focal class. Figure~\ref{fig:erq1_effect_heatmap} presents the percentage increases for significant associations as a green-hue heatmap. For binary IVs, percentage increases are obtained from the IRRs; for non-binary IVs, they are obtained from the doubling associations. For example, an IRR or doubling association of 1.10 corresponds to a 10\% increase. Here, $\boldsymbol{H1.1}$ uses a binary IV, whereas $\boldsymbol{H1.2}$ uses a non-binary IV. Light-gray cells indicate associations that are not significant, while dark-gray cells indicate class-level smell types not analyzed under $\boldsymbol{H1.2}$ because their focal-class counts are binary, as explained in Section~\ref{sec:rq_hypo}.

Across $\boldsymbol{H1.1}$ and $\boldsymbol{H1.2}$, 10 smell-specific associations are significant for \texttt{ChF}, compared with 7 for \texttt{ChS}. DP, FE, IT, and GC are significant for both \texttt{ChF} and \texttt{ChS} in at least one hypothesis, while SG and TB are associated with \texttt{ChF} but not \texttt{ChS}. Among smell types significantly associated with both \texttt{ChF} and \texttt{ChS}, the \texttt{ChS} effect sizes are 3--15 percentage points larger than the corresponding \texttt{ChF} effect sizes; the largest difference is for FE under $\boldsymbol{H1.1}$ (29\% for \texttt{ChS} versus 14\% for \texttt{ChF}). One notable contrast between effect size and prevalence is TB under $\boldsymbol{H1.1}$: it has the largest \texttt{ChF} effect size (28\%) but is also the rarest smell, occurring in only 0.08\% of classes.

\vspace{1mm}
\noindent
\fbox{%
  \parbox{\columnwidth}{%
    \hspace{3pt}%
    \parbox{0.95\columnwidth}{%
      \vspace{1mm}
      \textbf{RQ1 \& ERQ1 Summary:}
      At the aggregate level, smell number and variety in the focal class are significantly associated with higher change frequency, but not change size. At the smell-type level, certain smell types are significantly associated with change frequency, change size, or both.
      \vspace{1mm}
    }%
    \hspace{3pt}%
  }%
}

\subsection{RQ2 and ERQ2: Code Smells in Efferent Neighbors}
\label{sec:results_rq2_erq2}

Similar to RQ1 and ERQ1, we first characterize the prevalence of code smells in efferent neighbors. Because the focal class can be either smelly or non-smelly, we characterize this prevalence using three populations: all classes, smelly classes, and non-smelly classes. As shown in Table~\ref{tab:rq2_population}, of the 287,580 analyzed classes, 54,702 are smelly and 232,878 are non-smelly. The second column reports the number of classes with efferent neighbors, while the third column reports the number of classes with smelly efferent neighbors; both are reported for each population.

\begin{table}[h]
    \centering
    \caption{Class populations by efferent-neighbor status.}
    \renewcommand{\arraystretch}{1.5}
    \scriptsize
    \begin{tabular}{@{}lrr@{}}
        \hline
        \textbf{Population} & \makecell{\textbf{\#Classes with}\\\textbf{efferent neighbor}} & \makecell{\textbf{\#Classes with smelly}\\\textbf{efferent neighbor}} \\
        \hline
        All (287,580) & 247,400 & 191,018 \\
        Smelly (54,702) & 50,720 & 46,659 \\
        Non-smelly (232,878) & 196,680 & 144,359 \\
        \hline
    \end{tabular}
    \label{tab:rq2_population}
\end{table}

\begin{figure*}[tb]
    \centering
    \includegraphics[width=\textwidth]{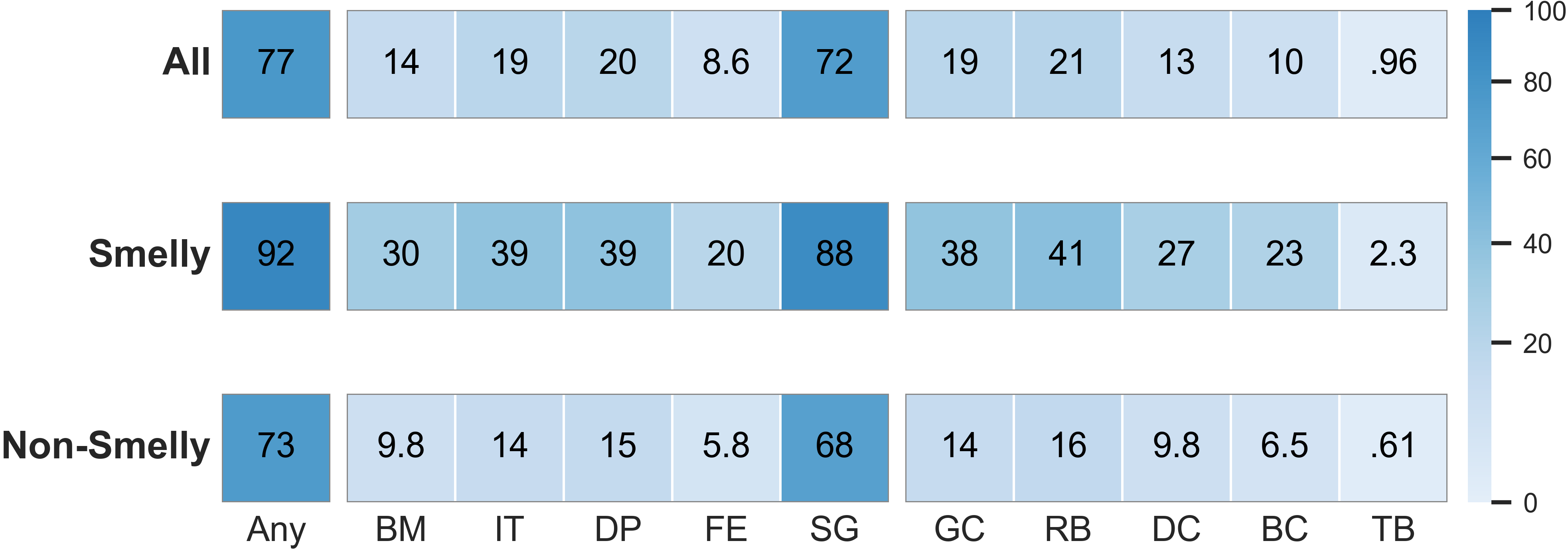}
    \caption{Prevalence of code smells in efferent neighbors.}
    \label{fig:erq2_efferent_prevalence}
\end{figure*}

Figure~\ref{fig:erq2_efferent_prevalence} shows the prevalence of code smells in efferent neighbors across the three populations. The ``Any'' category on the x-axis represents the prevalence of having at least one smelly efferent neighbor, calculated as \(\frac{\#\text{Classes with smelly efferent neighbor}}{\#\text{Classes with efferent neighbor}}\times 100\%\). The remaining categories show the prevalence of each specific code smell. For a code smell type \(CS\), prevalence was calculated as \(\frac{\#\text{Classes with smelly efferent neighbor}_{\text{[CS]}}}{\#\text{Classes with efferent neighbor}}\times 100\%\), where \textit{\#Classes with smelly efferent neighbor}\(_{\text{[CS]}}\) denotes the number of classes with at least one efferent neighbor exhibiting smell type \(CS\). These calculations were performed separately for the all, smelly, and non-smelly populations. We used \textit{\#Classes with efferent neighbor} as the denominator because classes without efferent neighbors have no opportunity to be exposed to code smells in efferent neighbors.

Smelly focal classes have higher efferent-neighbor prevalence than non-smelly focal classes for every smell type, by 1.69--25 percentage points. The largest gap is 25 percentage points for both IT and RB; for example, RB occurs in the efferent neighbors of 41\% of smelly classes versus 16\% of non-smelly classes. Overall, 77\% of classes with an efferent neighbor have at least one smelly efferent neighbor. This proportion is especially high among smelly focal classes (92\%), but remains high even among non-smelly focal classes (73\%).

For comparison, Figure~\ref{fig:erq1_focal_prevalence} shows that only 19\% of all analyzed classes exhibit at least one code smell themselves. Thus, code smells are substantially more prevalent in efferent neighbors than in focal classes. The same pattern holds for every individual smell type: efferent-neighbor prevalence exceeds focal-class prevalence by 0.88--68.1 percentage points. SG shows the largest difference (72\% in efferent neighbors versus 3.9\% in focal classes).

\vspace{1mm}
\noindent
\fbox{%
  \parbox{\columnwidth}{%
    \hspace{3pt}%
    \parbox{0.95\columnwidth}{%
      \vspace{1mm}
      \textbf{Finding 1:}
      Code smells are substantially more prevalent in efferent neighbors than in focal classes. Only 19\% of all analyzed focal classes are smelly, whereas among classes with at least one efferent neighbor, 77\% have at least one smelly efferent neighbor. Smelly efferent neighbors are common even among non-smelly focal classes (73\%), and even more prevalent among smelly focal classes (92\%).
      \vspace{1mm}
    }%
    \hspace{3pt}%
  }%
}

\subsubsection{RQ2: Do code smells in a class's efferent neighbors influence its stability?}

Table~\ref{tab:rq2_results} presents the RQ2 results. At the aggregate level, neither the presence nor the number of code smells in efferent neighbors is significantly associated with \texttt{ChF} or \texttt{ChS} in either the all-class or non-smelly-class population. Smell variety, however, shows significant associations: for all classes, doubling the variety is associated with 1.7\% higher \texttt{ChF} and 5.3\% higher \texttt{ChS}; among non-smelly classes, it is associated with 6.3\% higher \texttt{ChS}, while the association with \texttt{ChF} is not significant.

\begin{table*}[tb]
    \centering
    \caption{RQ2 results.}
    \small
    \renewcommand{\arraystretch}{1.25}
    \begin{tabular}{lrrrr}
        \hline
        \textbf{Hypothesis} & \textbf{IRR} & \textbf{AME} & \makecell{\textbf{Doubling}\\\textbf{association}} & $\boldsymbol{p}$ \\
        \hline
        $\boldsymbol{H2.1^{\text{\texttt{ChF}}}}$ & 0.947 & -0.036 & -- & 1.000 \\
        $\boldsymbol{H2.1^{\text{\texttt{ChS}}}}$ & 0.913 & -1.083 & -- & 1.000 \\
        $\boldsymbol{H2.2^{\text{\texttt{ChF}}}}$ & 0.986 & -0.002 & 0.990 & 1.000 \\
        $\boldsymbol{H2.2^{\text{\texttt{ChS}}}}$ & 0.971 & -0.073 & 0.980 & 1.000 \\
        $\boldsymbol{H2.3^{\text{\texttt{ChF}}}}$ & 1.025 & 0.006 & 1.017 & \textbf{0.005} \\
        $\boldsymbol{H2.3^{\text{\texttt{ChS}}}}$ & 1.078 & 0.319 & 1.053 & $\boldsymbol{3.01\times10^{-7}}$ \\
        $\boldsymbol{H2.4^{\text{\texttt{ChF}}}}$ & 0.963 & -0.017 & -- & 1.000 \\
        $\boldsymbol{H2.4^{\text{\texttt{ChS}}}}$ & 0.929 & -0.500 & -- & 1.000 \\
        $\boldsymbol{H2.5^{\text{\texttt{ChF}}}}$ & 0.984 & -0.002 & 0.989 & 1.000 \\
        $\boldsymbol{H2.5^{\text{\texttt{ChS}}}}$ & 0.978 & -0.053 & 0.985 & 1.000 \\
        $\boldsymbol{H2.6^{\text{\texttt{ChF}}}}$ & 1.019 & 0.004 & 1.013 & 0.076 \\
        $\boldsymbol{H2.6^{\text{\texttt{ChS}}}}$ & 1.092 & 0.298 & 1.063 & $\boldsymbol{3.01\times10^{-7}}$ \\
        \hline
    \end{tabular}
    \label{tab:rq2_results}
\end{table*}

\subsubsection{ERQ2: Which specific code smell types in a class's efferent neighbors influence its stability?}

\begin{figure*}[tb]
    \centering
    \includegraphics[width=\textwidth]{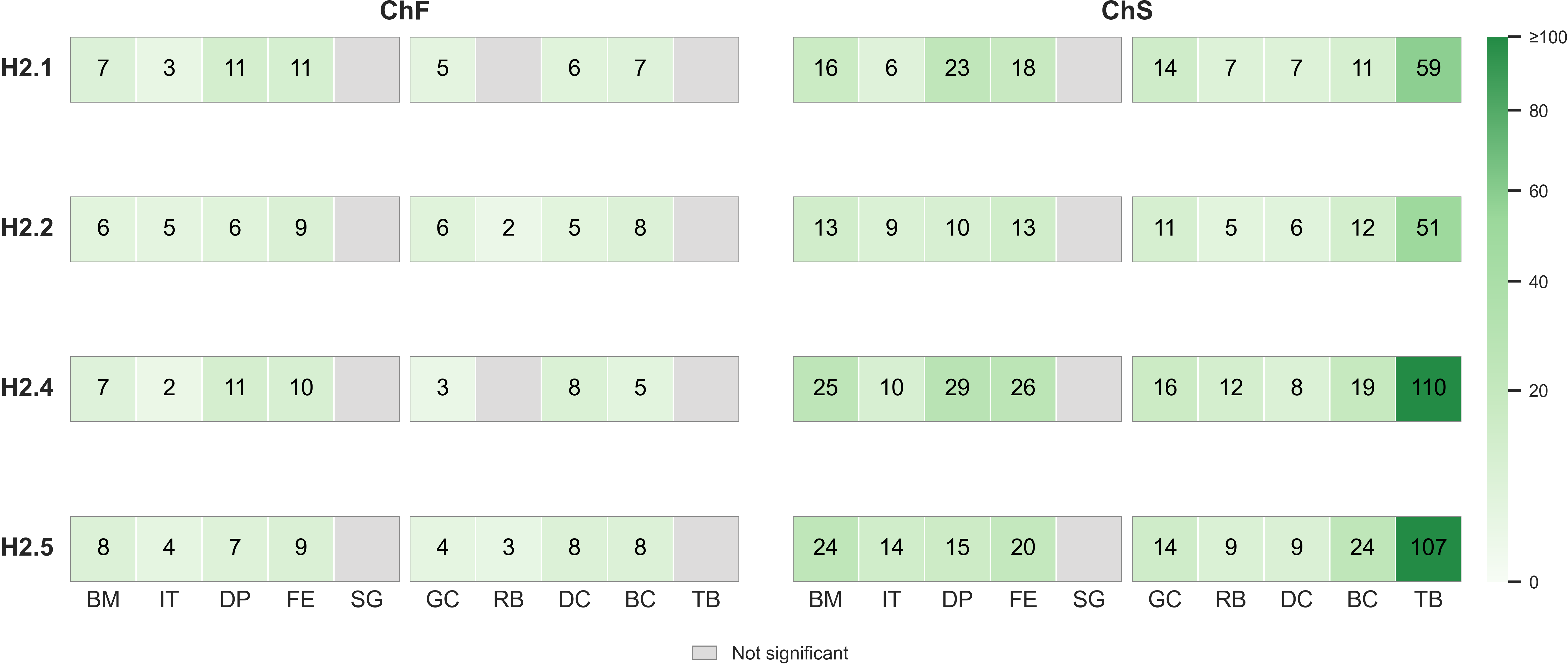}
    \caption{Percentage increases in \texttt{ChF} and \texttt{ChS} associated with code smells in efferent neighbors.}
    \label{fig:erq2_effect_heatmap}
\end{figure*}

ERQ2 examines smell-specific associations in efferent neighbors. Figure~\ref{fig:erq2_effect_heatmap} presents the percentage increases for significant associations. Seven smell types (BM, IT, DP, FE, GC, DC, and BC) are significantly associated with both \texttt{ChF} and \texttt{ChS} under all four hypotheses, showing a consistent pattern across the all and non-smelly populations. SG is the only smell with no significant association with either \texttt{ChF} or \texttt{ChS} under any of the four hypotheses. TB is significantly associated with \texttt{ChS} under all four hypotheses, but not with \texttt{ChF}. Its \texttt{ChS} effect sizes are 59\%, 51\%, 110\%, and 107\% under $\boldsymbol{H2.1}$, $\boldsymbol{H2.2}$, $\boldsymbol{H2.4}$, and $\boldsymbol{H2.5}$, respectively; these are the four largest effect sizes in the ERQ2 analysis. RB is also significantly associated with \texttt{ChS} under all four hypotheses. For \texttt{ChF}, the presence of RB in efferent neighbors is not significant in either the all-class or non-smelly-class population, whereas the number of RB instances is significant in both populations (2\% and 3\%, respectively).

Among smell types significantly associated with both \texttt{ChF} and \texttt{ChS}, the \texttt{ChS} effect sizes are 0--18 percentage points larger than the corresponding \texttt{ChF} effect sizes; the largest difference is under $\boldsymbol{H2.4}$ for both DP (29\% for \texttt{ChS} versus 11\% for \texttt{ChF}) and BM (25\% versus 7\%). The \texttt{ChS} effect sizes are also 1--56 percentage points larger in the non-smelly population than in the all-class population; the largest difference is for the number of TB instances in efferent neighbors (107\% under $\boldsymbol{H2.5}$ versus 51\% under $\boldsymbol{H2.2}$).

These smell-specific results contrast with the RQ2 results for aggregated smell measures in Table~\ref{tab:rq2_results}, where smell presence and number are not significantly associated with either \texttt{ChF} or \texttt{ChS} in either the all-class or non-smelly-class population. One possible contributor is SG: it is the most frequently occurring smell in efferent neighbors (72\% overall, 88\% among smelly classes, and 68\% among non-smelly classes), yet it shows no significant association with either \texttt{ChF} or \texttt{ChS} under any of the four hypotheses. Because the aggregate presence and number measures combine all smell types, the very high prevalence of SG may cause the aggregate measures to be dominated by SG and therefore reduce their ability to reflect associations from less prevalent smell types. This suggests that associations with stability for specific smell types in efferent neighbors can be obscured by aggregate measures.

Comparing these results with the focal-class results in Figure~\ref{fig:erq1_effect_heatmap}, DP, FE, GC, and IT show significant associations when occurring in the focal class and in efferent neighbors. In contrast, BM, DC, BC, and RB show no significant association in the focal class but show significant associations when occurring in efferent neighbors. SG shows the opposite pattern: its presence and number are associated with \texttt{ChF} when it occurs in the focal class, but are not significantly associated with either \texttt{ChF} or \texttt{ChS} when it occurs in efferent neighbors.

\vspace{1mm}
\noindent
\fbox{%
  \parbox{\columnwidth}{%
    \hspace{3pt}%
    \parbox{0.95\columnwidth}{%
      \vspace{1mm}
      \textbf{Finding 2:}
      Associations with stability depend on the location in which a specific smell occurs: some smell types show significant associations when they occur in focal classes and in efferent neighbors, whereas others show significant associations only when they occur in efferent neighbors.
      \vspace{1mm}
    }%
    \hspace{3pt}%
  }%
}

\vspace{1mm}
\noindent
\fbox{%
  \parbox{\columnwidth}{%
    \hspace{3pt}%
    \parbox{0.95\columnwidth}{%
      \vspace{1mm}
      \textbf{RQ2 \& ERQ2 Summary:}
      At the aggregate level, smell presence and number in efferent neighbors are not significantly associated with either change frequency or change size. At the smell-type level, however, the presence and number of seven smell types are consistently associated with both change frequency and change size, and this pattern also holds among non-smelly focal classes. Associations with stability also depend on the location in which a specific smell occurs.
      \vspace{1mm}
    }%
    \hspace{3pt}%
  }%
}

\subsection{RQ3: Does efferent code smell coupling influence a class's stability?}
\label{sec:results_rq3}

\begin{table*}[tb]
    \centering
    \caption{RQ3 results.}
    \small
    \renewcommand{\arraystretch}{1.25}
    \begin{tabular}{lrrrr}
        \hline
        \textbf{Hypothesis} & \textbf{IRR} & \textbf{AME} & \makecell{\textbf{Doubling}\\\textbf{association}} & $\boldsymbol{p}$ \\
        \hline
        $\boldsymbol{H3.1^{\text{\texttt{ChF}}}}$ & 1.031 & 0.020 & -- & \textbf{0.00078} \\
        $\boldsymbol{H3.1^{\text{\texttt{ChS}}}}$ & 0.991 & -0.101 & -- & 0.691 \\
        $\boldsymbol{H3.2^{\text{\texttt{ChF}}}}$ & 1.178 & 0.108 & -- & $\boldsymbol{3.36\times10^{-11}}$ \\
        $\boldsymbol{H3.2^{\text{\texttt{ChS}}}}$ & 1.083 & 0.924 & -- & \textbf{0.027} \\
        \hline
    \end{tabular}
    \label{tab:rq3_results}
\end{table*}

Table~\ref{tab:rq3_results} presents the RQ3 results. When the separate presence of code smells in the focal class and its efferent neighbors is not controlled for ($\boldsymbol{H3.1}$), efferent code smell coupling is significantly associated with \texttt{ChF}, but not with \texttt{ChS}. Classes with efferent code smell coupling have 3.1\% higher \texttt{ChF}. After controlling for the separate presence of code smells in the focal class and its efferent neighbors ($\boldsymbol{H3.2}$), efferent code smell coupling is significantly associated with both stability outcomes, corresponding to 17.8\% higher \texttt{ChF} and 8.3\% higher \texttt{ChS}.

\vspace{1mm}
\noindent
\fbox{%
  \parbox{\columnwidth}{%
    \hspace{3pt}%
    \parbox{0.95\columnwidth}{%
      \vspace{1mm}
      \textbf{RQ3 Summary:}
      Efferent code smell coupling shows an additional association with 17.8\% higher change frequency and 8.3\% higher change size beyond the separate associations of code smell presence in the focal class and in efferent neighbors.
      \vspace{1mm}
    }%
    \hspace{3pt}%
  }%
}

\subsection{RQ4 and ERQ4: Efferent Code Smell Interaction}
\label{sec:results_rq4_erq4}

\subsubsection{RQ4: Does efferent code smell interaction influence a class's stability?}

Table~\ref{tab:rq4_results} presents the RQ4 results. When the numbers of code smells in the focal class and its efferent neighbors are not controlled for, the presence of an efferent code smell interaction ($\boldsymbol{H4.1}$) is not significantly associated with either \texttt{ChF} or \texttt{ChS}. In contrast, the number of efferent code smell interactions ($\boldsymbol{H4.2}$) and interaction intensity ($\boldsymbol{H4.3}$) are significantly associated with both \texttt{ChF} and \texttt{ChS}. Doubling the number of interactions is associated with 1.1\% higher \texttt{ChF} and 1.2\% higher \texttt{ChS}, while doubling interaction intensity is associated with 0.6\% higher \texttt{ChF} and 1.0\% higher \texttt{ChS}.

After controlling for the numbers of code smells in the focal class and its efferent neighbors, interaction presence ($\boldsymbol{H4.4}$) is not significantly associated with either \texttt{ChF} or \texttt{ChS}. The number of interactions ($\boldsymbol{H4.5}$) and interaction intensity ($\boldsymbol{H4.6}$) are significantly associated with \texttt{ChS}, but not \texttt{ChF}. Doubling the number of interactions is associated with 2.9\% higher \texttt{ChS}, while doubling interaction intensity is associated with 2.3\% higher \texttt{ChS}. Thus, after accounting for the overall amount of code smells in the two locations, the aggregate number and intensity of efferent code smell interactions remain associated with change size, but not change frequency.

\begin{table*}[tb]
    \centering
    \caption{RQ4 results.}
    \small
    \renewcommand{\arraystretch}{1.25}
    \begin{tabular}{lrrrr}
        \hline
        \textbf{Hypothesis} & \textbf{IRR} & \textbf{AME} & \makecell{\textbf{Doubling}\\\textbf{association}} & $\boldsymbol{p}$ \\
        \hline
        $\boldsymbol{H4.1^{\text{\texttt{ChF}}}}$ & 1.015 & 0.009 & -- & 0.099 \\
        $\boldsymbol{H4.1^{\text{\texttt{ChS}}}}$ & 1.024 & 0.270 & -- & 0.142 \\
        $\boldsymbol{H4.2^{\text{\texttt{ChF}}}}$ & 1.017 & 0.007 & 1.011 & $\boldsymbol{2.60\times10^{-5}}$ \\
        $\boldsymbol{H4.2^{\text{\texttt{ChS}}}}$ & 1.017 & 0.113 & 1.012 & \textbf{0.020} \\
        $\boldsymbol{H4.3^{\text{\texttt{ChF}}}}$ & 1.009 & 0.004 & 1.006 & \textbf{0.004} \\
        $\boldsymbol{H4.3^{\text{\texttt{ChS}}}}$ & 1.015 & 0.098 & 1.010 & \textbf{0.019} \\
        $\boldsymbol{H4.4^{\text{\texttt{ChF}}}}$ & 0.939 & -0.040 & -- & 1.000 \\
        $\boldsymbol{H4.4^{\text{\texttt{ChS}}}}$ & 1.051 & 0.568 & -- & 0.052 \\
        $\boldsymbol{H4.5^{\text{\texttt{ChF}}}}$ & 0.992 & -0.003 & 0.995 & 1.000 \\
        $\boldsymbol{H4.5^{\text{\texttt{ChS}}}}$ & 1.042 & 0.275 & 1.029 & $\boldsymbol{4.73\times10^{-4}}$ \\
        $\boldsymbol{H4.6^{\text{\texttt{ChF}}}}$ & 0.985 & -0.006 & 0.989 & 1.000 \\
        $\boldsymbol{H4.6^{\text{\texttt{ChS}}}}$ & 1.034 & 0.218 & 1.023 & $\boldsymbol{4.73\times10^{-4}}$ \\
        \hline
    \end{tabular}
    \label{tab:rq4_results}
\end{table*}

\subsubsection{ERQ4: Which smell-type pairs influence a class's stability through efferent code smell interaction?}

\begin{figure*}[tbh]
    \centering
    \includegraphics[width=0.8\textwidth]{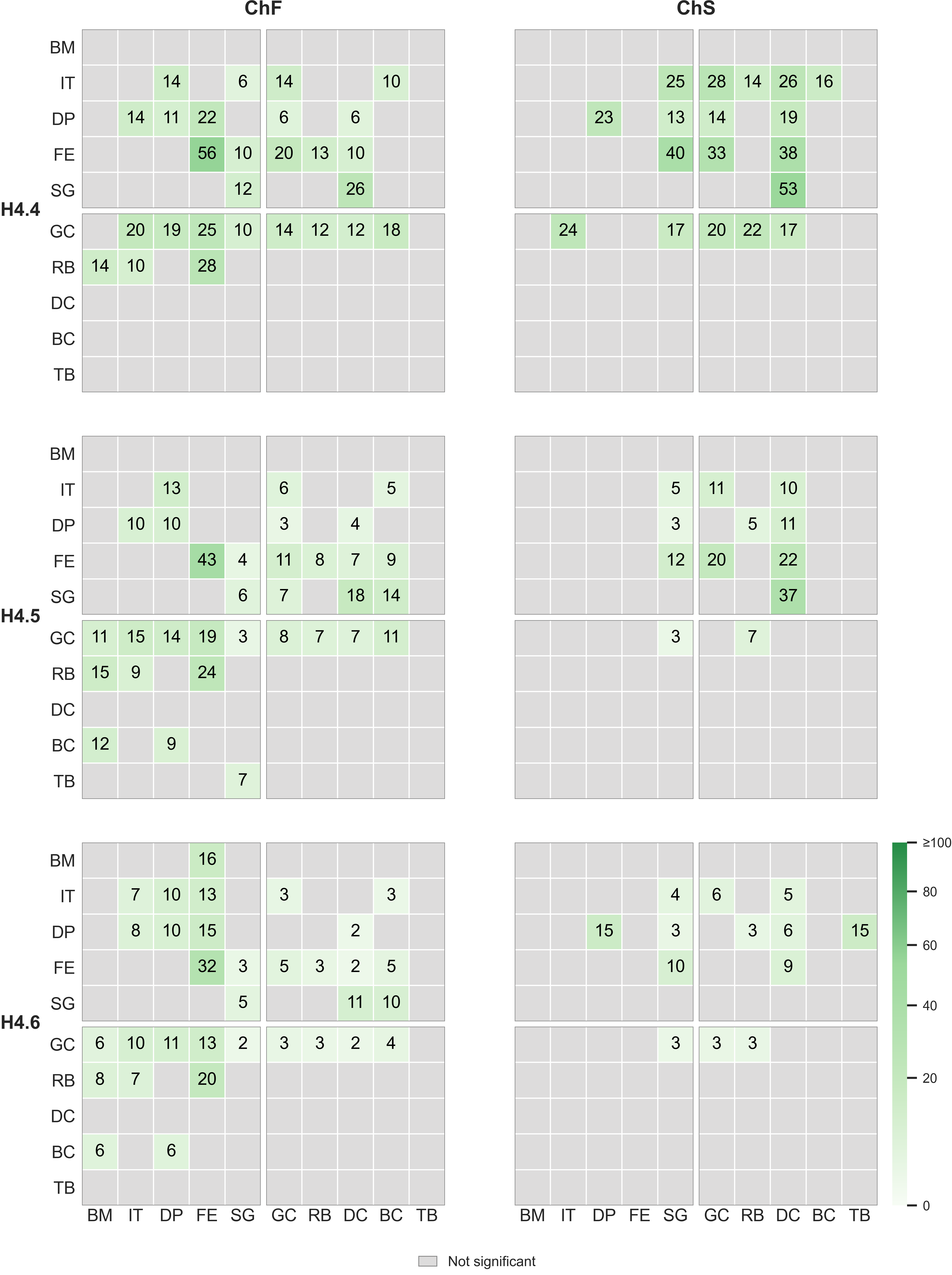}
    \caption{Percentage increases in \texttt{ChF} and \texttt{ChS} associated with efferent code smell interactions by directional smell-type pair.}
    \label{fig:erq4_pair_effect_heatmap}
\end{figure*}

ERQ4 examines which directional smell-type pairs are associated with stability through efferent code smell interaction. Figure~\ref{fig:erq4_pair_effect_heatmap} presents the percentage increases for significant associations under $\boldsymbol{H4.4}$--$\boldsymbol{H4.6}$ for \texttt{ChF} and \texttt{ChS}. The y-axis represents $\mathrm{CS}_1$, the source smell type exhibited by the source artifact in the focal class, while the x-axis represents $\mathrm{CS}_2$, the target smell type exhibited by the target artifact in the efferent neighbor. Each cell represents the directional smell-type pair $\mathrm{CS}_1\rightarrow\mathrm{CS}_2$. On each axis, the first 5 smell types (BM, IT, DP, FE, and SG) are method-level smells and the last 5 (GC, RB, DC, BC, and TB) are class-level smells, dividing each panel into method-to-method, method-to-class, class-to-method, and class-to-class quadrants. Green cells indicate significant associations, while gray cells indicate associations that are not significant.

Across the six panels, 42 of the 100 directional smell-type pairs are significantly associated with \texttt{ChF} or \texttt{ChS} under at least one hypothesis, while 58 pairs show no significant association in any panel. Under each of \texttt{ChF} and \texttt{ChS}, the significant pairs are largely similar across the three hypotheses. For \texttt{ChF}, 27, 32, and 33 pairs are significant under $\boldsymbol{H4.4}$, $\boldsymbol{H4.5}$, and $\boldsymbol{H4.6}$, respectively, and 24 are significant under all three hypotheses. For \texttt{ChS}, the corresponding counts are 18, 12, and 13, with 9 pairs significant under all three hypotheses. Six pairs are significant in all six panels: DP$\rightarrow$DC, FE$\rightarrow$DC, FE$\rightarrow$SG, GC$\rightarrow$RB, GC$\rightarrow$SG, and IT$\rightarrow$GC. This contrasts with the aggregate RQ4 results for \texttt{ChF} in Table~\ref{tab:rq4_results}: none of $\boldsymbol{H4.4}$--$\boldsymbol{H4.6}$ is significant at the aggregate level, whereas 27--33 of the 100 smell-type pairs are significant across the three hypotheses. This indicates that associations with change frequency can be obscured when interactions are aggregated across smell-type pairs.

For \texttt{ChS}, significant associations are concentrated in interactions from method-level smells to class-level smells. This quadrant contains 9 of 18 significant cells under $\boldsymbol{H4.4}$, 7 of 12 under $\boldsymbol{H4.5}$, and 6 of 13 under $\boldsymbol{H4.6}$, accounting for 22 of the 43 significant \texttt{ChS} cells across the three panels. Class-to-class interactions are sparse for both \texttt{ChF} and \texttt{ChS} and are consistently the sparsest quadrant for \texttt{ChF}, with four significant cells under each hypothesis. Effect sizes also tend to be larger for \texttt{ChS}: among the 31 cases in which the same smell-type pair is significant for both \texttt{ChF} and \texttt{ChS} under the same hypothesis, the \texttt{ChS} effect size is larger in 27 cases (87\%). The largest difference is FE$\rightarrow$SG under $\boldsymbol{H4.4}$ (40\% for \texttt{ChS} versus 10\% for \texttt{ChF}). The largest individual effect sizes are FE$\rightarrow$FE for \texttt{ChF} under $\boldsymbol{H4.4}$ (56\%) and SG$\rightarrow$DC for \texttt{ChS} under $\boldsymbol{H4.4}$ (53\%).

Because smell-type pairs are directional, we next examined whether the observed significance pattern differs when the source and target roles are reversed. Excluding the 10 diagonal pairs leaves 90 directional pairs, corresponding to 45 combinations of two distinct smell types $\mathrm{CS}_1$ and $\mathrm{CS}_2$; for each combination, we compared $\mathrm{CS}_1\rightarrow\mathrm{CS}_2$ with $\mathrm{CS}_2\rightarrow\mathrm{CS}_1$, treating a direction as significant if it is significant under at least one of $\boldsymbol{H4.4}$--$\boldsymbol{H4.6}$. For \texttt{ChF}, 26 combinations have at least one significant direction; of these, 20 (77\%) are significant in only one direction and 6 (23\%) are significant in both. For \texttt{ChS}, 17 combinations have at least one significant direction; of these, 16 (94\%) are significant in only one direction, while only the combination of IT and GC is significant in both directions.

\vspace{1mm}
\noindent
\fbox{%
  \parbox{\columnwidth}{%
    \hspace{3pt}%
    \parbox{0.95\columnwidth}{%
      \vspace{1mm}
      \textbf{Finding 3:}
      Significant associations between efferent code smell interaction and stability are frequently direction-specific. For combinations of two smell types $\mathrm{CS}_1$ and $\mathrm{CS}_2$ with at least one significant direction ($\mathrm{CS}_1\rightarrow\mathrm{CS}_2$ or $\mathrm{CS}_2\rightarrow\mathrm{CS}_1$), 77\% for \texttt{ChF} and 94\% for \texttt{ChS} are significant in only one direction.
      \vspace{1mm}
    }%
    \hspace{3pt}%
  }%
}

To further characterize these source and target roles, Tables~\ref{tab:erq4_source_target_chf} and~\ref{tab:erq4_source_target_chs} summarize, for each smell, how many different smell types it is significantly paired with when acting as the source and when acting as the target. For each smell type $CS$ and outcome $O \in \{\texttt{ChF}, \texttt{ChS}\}$, let $S(CS,O)=\{\mathrm{CS}_2: CS\rightarrow \mathrm{CS}_2 \text{ is significant for } O \text{ under at least one of } \boldsymbol{H4.4}\text{--}\boldsymbol{H4.6}\}$ and $T(CS,O)=\{\mathrm{CS}_1: \mathrm{CS}_1\rightarrow CS \text{ is significant for } O \text{ under at least one of } \boldsymbol{H4.4}\text{--}\boldsymbol{H4.6}\}$. Source count is $|S(CS,O)|$, the number of smell types significantly paired with $CS$ when $CS$ acts as the source. Target count is $|T(CS,O)|$, the number of smell types significantly paired with $CS$ when $CS$ acts as the target. Difference is calculated as $|T(CS,O)|-|S(CS,O)|$; a positive value means that $CS$ is significantly paired with more smell types when acting as the target, while a negative value means that it is significantly paired with more smell types when acting as the source. Overlap is calculated as $\frac{|S(CS,O)\cap T(CS,O)|}{|S(CS,O)\cup T(CS,O)|}$; values closer to 1 mean that the same smell types tend to be significantly paired with $CS$ in both roles. Overlap is undefined and shown as ``--'' when $S(CS,O)\cup T(CS,O)$ is empty.

\begin{table}[tb]
    \centering
    \caption{Source and target role patterns of smells for \texttt{ChF}.}
    \small
    \renewcommand{\arraystretch}{1.25}
    \setlength{\tabcolsep}{4pt}
    \begin{tabular}{lrrrr}
        \hline
        \textbf{Smell} & \textbf{Source count} & \textbf{Target count} & \textbf{Difference} & \textbf{Overlap} \\
        \hline
        BM & 1 & 3 & +2 & 0.00 \\
        IT & 6 & 4 & -2 & 0.43 \\
        DP & 5 & 4 & -1 & 0.50 \\
        FE & 6 & 6 & 0 & 0.33 \\
        SG & 4 & 5 & +1 & 0.29 \\
        GC & 9 & 5 & -4 & 0.56 \\
        RB & 3 & 2 & -1 & 0.25 \\
        DC & 0 & 4 & +4 & 0.00 \\
        BC & 2 & 4 & +2 & 0.00 \\
        TB & 1 & 0 & -1 & 0.00 \\
        \hline
    \end{tabular}
    \label{tab:erq4_source_target_chf}
\end{table}

\begin{table}[tb]
    \centering
    \caption{Source and target role patterns of smells for \texttt{ChS}.}
    \small
    \renewcommand{\arraystretch}{1.25}
    \setlength{\tabcolsep}{4pt}
    \begin{tabular}{lrrrr}
        \hline
        \textbf{Smell} & \textbf{Source count} & \textbf{Target count} & \textbf{Difference} & \textbf{Overlap} \\
        \hline
        BM & 0 & 0 & 0 & -- \\
        IT & 5 & 1 & -4 & 0.20 \\
        DP & 6 & 1 & -5 & 0.17 \\
        FE & 3 & 0 & -3 & 0.00 \\
        SG & 1 & 4 & +3 & 0.00 \\
        GC & 5 & 4 & -1 & 0.29 \\
        RB & 0 & 3 & +3 & 0.00 \\
        DC & 0 & 5 & +5 & 0.00 \\
        BC & 0 & 1 & +1 & 0.00 \\
        TB & 0 & 1 & +1 & 0.00 \\
        \hline
    \end{tabular}
    \label{tab:erq4_source_target_chs}
\end{table}

For \texttt{ChF}, GC has the largest source count: when GC acts as the source, it is significantly paired with 9 different smell types, whereas when it acts as the target, it is significantly paired with 5. DC shows the strongest contrast in the opposite direction: it is significantly paired with 4 smell types when acting as the target but with none when acting as the source, giving the largest positive difference ($+4$). FE is significantly paired with 6 smell types in each role, but its overlap is only 0.33, indicating that the sets of smell types paired with FE differ substantially between the two roles. GC, DP, and IT have the highest overlaps (0.56, 0.50, and 0.43, respectively), whereas several other smells have little or no overlap.

The source--target distinction is even clearer for \texttt{ChS}. DP, IT, and FE are significantly paired with more smell types when acting as the source than when acting as the target, with differences of $-5$, $-4$, and $-3$, respectively. In contrast, DC, SG, and RB are significantly paired with more smell types when acting as the target, with differences of $+5$, $+3$, and $+3$. Overlap is low overall, reaching only 0.29 for GC, 0.20 for IT, and 0.17 for DP, and zero for the remaining smells with at least one significant association. This low overlap is expected given the directionality analysis above, where 94\% of the \texttt{ChS} combinations with at least one significant direction are significant in only one direction.

\vspace{1mm}
\noindent
\fbox{%
  \parbox{\columnwidth}{%
    \hspace{3pt}%
    \parbox{0.95\columnwidth}{%
      \vspace{1mm}
      \textbf{Finding 4:}
      Certain smell types tend to show significant associations when interacting with other smells as the source smell type, whereas others tend to show significant associations as the target smell type.
      \vspace{1mm}
    }%
    \hspace{3pt}%
  }%
}

Comparing ERQ4 with ERQ1 (Figure~\ref{fig:erq1_effect_heatmap}) and ERQ2 (Figure~\ref{fig:erq2_effect_heatmap}), BM, BC, and RB show no significant association when they occur in the focal class in ERQ1, yet as source smell types in ERQ4 they are significantly paired with 1, 2, and 3 different smell types, respectively, for \texttt{ChF}. Conversely, the presence and number of SG in efferent neighbors are not significantly associated with either \texttt{ChF} or \texttt{ChS} in ERQ2, yet as a target smell type in ERQ4, SG is significantly paired with 5 different smell types for \texttt{ChF} and 4 for \texttt{ChS}. FE$\rightarrow$SG and GC$\rightarrow$SG are significant in all six ERQ4 panels.

\vspace{1mm}
\noindent
\fbox{%
  \parbox{\columnwidth}{%
    \hspace{3pt}%
    \parbox{0.95\columnwidth}{%
      \vspace{1mm}
      \textbf{Finding 5:}
      Efferent code smell interaction reveals associations not observed when smell types are analyzed at one location alone: some smell types with no significant association in the focal class show significant associations when appearing as source smell types in specific smell-type pairs, while some with no significant association in efferent neighbors show significant associations when appearing as target smell types in specific smell-type pairs.
      \vspace{1mm}
    }%
    \hspace{3pt}%
  }%
}

We finally examined the prevalence of the significant smell-type pairs to provide practical context for the statistically significant associations. For a smell-type pair $\mathrm{CS}_1\rightarrow\mathrm{CS}_2$, prevalence was calculated as \(\frac{\#\text{Classes with efferent code smell interaction}_{[\mathrm{CS}_1\rightarrow\mathrm{CS}_2]}}{\#\text{Classes with efferent code smell interaction}}\times 100\%\). Here, \textit{\#Classes with efferent code smell interaction}$_{[\mathrm{CS}_1\rightarrow\mathrm{CS}_2]}$ denotes the number of classes exhibiting at least one efferent code smell interaction of that specific smell-type pair, while \textit{\#Classes with efferent code smell interaction} denotes the number of classes with at least one efferent code smell interaction of any smell-type pair. The denominator therefore consisted of 39,831 classes (13.9\% of the 287,580 analyzed classes). We used \textit{\#Classes with efferent code smell interaction} as the denominator because only these classes can exhibit efferent code smell interaction of a specific smell-type pair, thereby focusing the prevalence calculation on the relevant population.

\begin{figure*}[tb]
    \centering
    \includegraphics[width=0.8\textwidth]{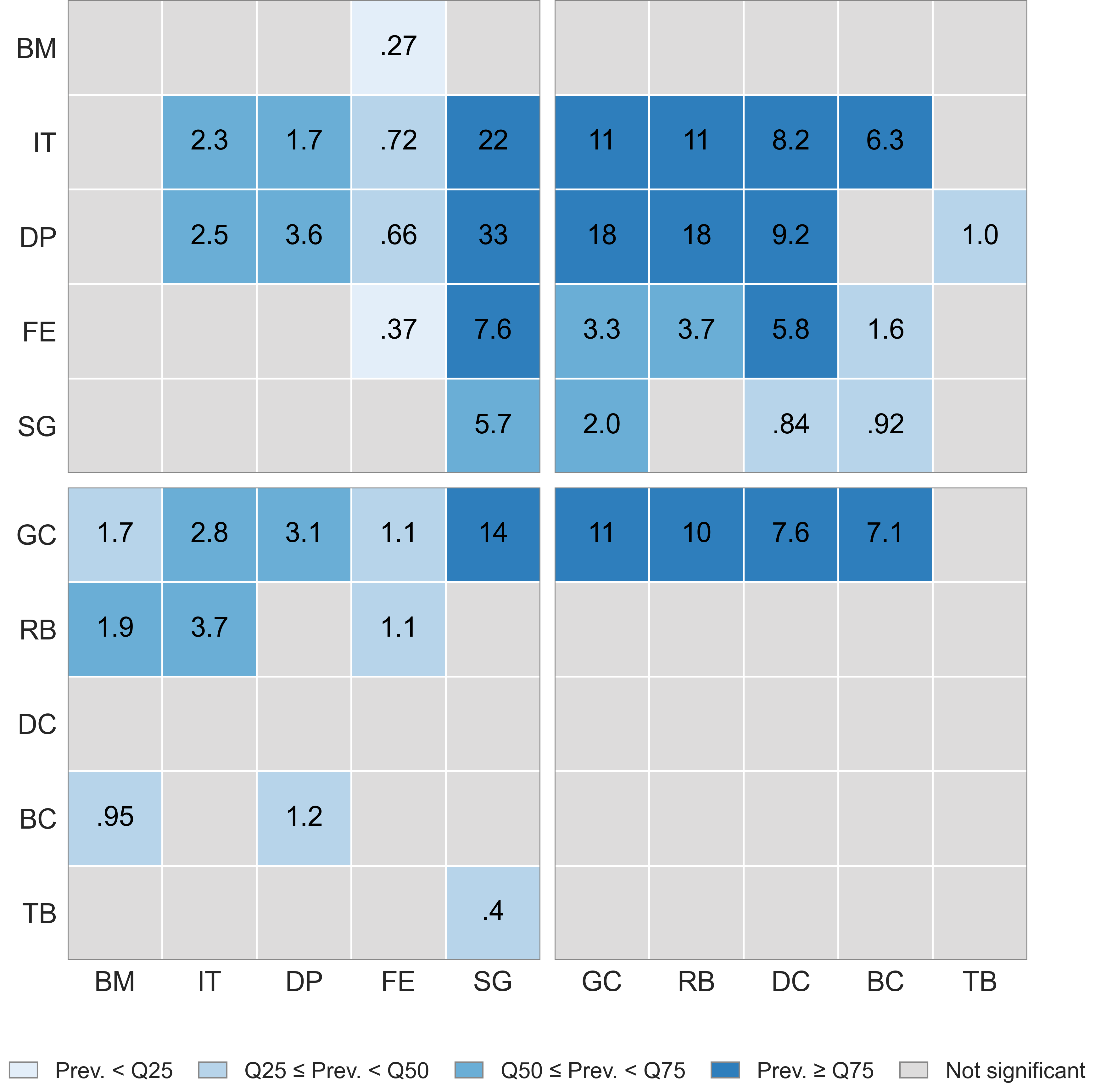}
    \caption{Prevalence of smell-type pairs with at least one significant association in ERQ4, grouped by prevalence quartile.}
    \label{fig:erq4_pair_prevalence}
\end{figure*}

Because there are 100 possible directional smell-type pairs, many of the raw prevalence values are concentrated at the low end. Using the raw percentages directly as a continuous color scale would therefore make many cells appear visually similar and make differences difficult to interpret. Instead of interpreting the raw prevalence values in isolation, we assessed relative prevalence by examining where each significant smell-type pair ranks among all 100 smell-type pairs. We therefore ranked all 100 pairs by their prevalence and used the 25th, 50th, and 75th percentile thresholds ($Q_{25}=0.38\%$, $Q_{50}=1.67\%$, and $Q_{75}=5.71\%$) to classify them into four prevalence quartiles. Figure~\ref{fig:erq4_pair_prevalence} displays the prevalence of pairs that are significant in at least one of the six panels in Figure~\ref{fig:erq4_pair_effect_heatmap}. The value in each cell is its raw prevalence, while the blue hue represents its prevalence quartile relative to all 100 pairs; gray cells are not significant in any of the six panels.

Most of the significant smell-type pairs are comparatively prevalent within the candidate pair space. Of the 42 pairs significant in at least one panel, 28 (67\%) are at or above the median prevalence ($Q_{50}$), including 16 (38\%) at or above $Q_{75}$. Only two significant pairs are below $Q_{25}$. Moreover, all six pairs that are significant across all six ERQ4 panels are at or above $Q_{75}$, with prevalence ranging from 5.8\% to 14\%.

\vspace{1mm}
\noindent
\fbox{%
  \parbox{\columnwidth}{%
    \hspace{3pt}%
    \parbox{0.95\columnwidth}{%
      \vspace{1mm}
      \textbf{Finding 6:}
      Smell-type pairs with significant associations are comparatively prevalent: 67\% are at or above the median prevalence, and 38\% are at or above the 75th percentile ($Q_{75}$).
      \vspace{1mm}
    }%
    \hspace{3pt}%
  }%
}

\vspace{1mm}
\noindent
\fbox{%
  \parbox{\columnwidth}{%
    \hspace{3pt}%
    \parbox{0.95\columnwidth}{%
      \vspace{1mm}
      \textbf{RQ4 \& ERQ4 Summary:}
      At the aggregate level, after controlling for the number of code smells, the number and intensity of efferent code smell interactions are significantly associated with change size but not change frequency. At the level of individual smell-type pairs, significant associations are frequently direction-specific and also depend on whether a smell interacts as the source or target smell type. Considering efferent code smell interactions also reveals associations with stability that are not observed when the corresponding smell types are analyzed at one location alone.
      \vspace{1mm}
    }%
    \hspace{3pt}%
  }%
}

\section{Discussion} \label{sec:discussion}

Our findings have several implications for how code smells are inspected and prioritized in maintenance, as well as for future research.

\subsection{Implication 1: Inspect Efferent Neighbors, Not Just the Focal Class}

Finding~1 shows that code smells are substantially more prevalent in efferent neighbors than in focal classes, and the prevalence of smells in efferent neighbors remains relatively high even among non-smelly focal classes. Finding~2 further shows that the association of a smell with stability depends on its location: several smell types that show no significant association when they occur in the focal class show significant associations when they occur in efferent neighbors. Therefore, focusing only on the class being maintained (i.e., the focal class) may overlook code smells in its efferent neighbors that are also potentially harmful to its stability.

A practical response is to make code smell reports dependency-aware. In addition to reporting that a focal class contains a particular smell, a tool could report which of its efferent neighbors are smelly and which smell types they contain. This is particularly useful for a non-smelly focal class, where a conventional smell report may provide little reason to inspect the class even though it depends on smelly neighbors that may be adverse to the class's stability. Such context would not change whether an artifact is classified as smelly; instead, it would enrich the report with information that can guide developers toward related artifacts that may warrant inspection during maintenance.

\subsection{Implication 2: Inspect Code Smell Interactions, Not Just Smells at Each Location}

RQ3 shows that, after accounting for code smell presence in the focal class and in its efferent neighbors separately, efferent code smell coupling still shows an additional association with 17.8\% higher change frequency and 8.3\% higher change size. One possible explanation is that some of this additional association reflects cases in which smells in the two locations directly interact through static dependencies.

Finding~5 provides further evidence consistent with this explanation: considering smells at the focal and efferent-neighbor locations separately can miss associations that become apparent when they form an efferent code smell interaction. For example, BM, BC, and RB show no significant association when they occur in the focal class, yet they show significant associations when appearing as source smell types in specific smell-type pairs. Conversely, SG shows no significant association in efferent neighbors, yet it shows significant associations when appearing as a target smell type in specific smell-type pairs.

Therefore, after identifying smells in a class and its efferent neighbors, practitioners should also examine whether those smells are directly connected through static dependencies and form an efferent code smell interaction. A dependency-aware smell report could make this relationship explicit by showing not only which smells occur on each side of a dependency, but also which of them interact. This can help developers identify cases that may warrant closer inspection even when the individual smells appear less concerning when considered separately.

\subsection{Implication 3: Prioritize Specific Smell-Type Pairs}

The ERQ4 results show why practitioners should consider specific smell-type pairs rather than treating all pairs uniformly. For \texttt{ChF}, none of the aggregate interaction measures under $\boldsymbol{H4.4}$--$\boldsymbol{H4.6}$ in RQ4 is significant, whereas 27--33 of the 100 smell-type pairs are significant across the corresponding ERQ4 analyses. Finding~3 further shows that direction matters: among code smell combinations $\mathrm{CS}_1$ and $\mathrm{CS}_2$ with at least one significant direction ($\mathrm{CS}_1\rightarrow\mathrm{CS}_2$ or $\mathrm{CS}_2\rightarrow\mathrm{CS}_1$), 77\% for \texttt{ChF} and 94\% for \texttt{ChS} are significant in only one direction. Thus, ignoring direction or aggregating interactions across smell-type pairs can obscure information that is useful for maintenance decisions.

Our ERQ4 effect-size results (Figure~\ref{fig:erq4_pair_effect_heatmap}) and prevalence results (Figure~\ref{fig:erq4_pair_prevalence}) can help practitioners and tools prioritize smell-type pairs for inspection or refactoring attention by identifying those that show stronger associations or occur relatively frequently.

\subsection{Implication 4: For Target-Dominant Smells, Inspect \underline{A}fferent Neighbors Rather Than \underline{E}fferent Neighbors}

Finding~4 shows that some smell types tend to be target-dominant, meaning that they exhibit significant associations with more smell types when appearing as a target smell type than as a source smell type. For \texttt{ChS}, Table~\ref{tab:erq4_source_target_chs} shows that DC, SG, and RB exhibit a clear target-dominant pattern. This distinction changes where practitioners may need to inspect after detecting such a smell. In our analysis, the source smell is located in the focal class whose stability is modeled, whereas the target smell is located in its efferent neighbor. Therefore, when a selected class contains a target-dominant smell, the classes that may warrant closer inspection are often the classes that depend on that selected class. We refer to these as its \emph{afferent neighbors}: afferent is the antonym of efferent, so the afferent neighbors of a class are the classes that depend on it, whereas its efferent neighbors are the classes that it depends on.

When a target-dominant smell is detected in a selected class, a tool could therefore help developers inspect the selected class's afferent neighbors and identify whether they form efferent code smell interactions in which the afferent neighbor contains the source smell and the selected class contains the target smell. The purpose of this approach is to quickly identify candidate classes whose stability may be affected by the selected smelly class. This differs from the efferent-neighbor inspection described in Implication~1, where inspecting a selected class's efferent neighbors helps practitioners assess factors that may affect the stability of the selected class itself.

\subsection{Implication 5: Investigate Potential Mechanisms Behind Smell-Type Pair Associations}

Ripple effects provide a general explanation for why efferent code smell interactions can be associated with lower class stability: changes to a smell in an efferent neighbor may propagate through the dependency to a smell in the focal class. However, Figure~\ref{fig:erq4_pair_effect_heatmap} shows substantial variation in effect sizes across smell-type pairs, suggesting that the mechanisms behind particular pairs may be more specific. For example, under $\boldsymbol{H4.4}$, the presence of SG$\rightarrow$DC is associated with 53\% higher \texttt{ChS}, the largest ChS effect size observed among the significant smell-type pairs. Shotgun Surgery (SG) characterizes a method that is invoked by many methods across many classes, whereas Data Class (DC) characterizes a class that mainly stores data and exposes it through public fields or accessor methods. When an SG method in the focal class directly depends on a DC in an efferent neighbor, a change to the data structure or interface of the DC may require corresponding changes to the SG method. Because the SG method serves many callers, adapting it while preserving its role for those callers may require more extensive changes in the focal class.

This is one plausible mechanism, and similar explanations are not apparent for many of the other significant smell-type pairs. Smell definitions characterize the properties of the interacting artifacts, but they do not describe how those artifacts are structurally connected. We therefore suspect that dependency type can provide an important additional layer of explanation. For SG$\rightarrow$DC, for example, whether the SG method creates the DC, receives it as a parameter, returns it, or otherwise depends on it may imply different consequences for stability. Future research could therefore investigate whether the association of a smell-type pair with class stability also depends on the dependency type connecting the two smell instances.

\subsection{Implication 6: Consider Smells in Efferent Neighbors in Architectural-Smell Evolution Research}

Our findings may also inform future research on architectural-smell evolution. Prior studies have mainly examined how individual architectural smells evolve over time. For example, \citet{sas2019investigating} studied their growth rate and lifetime, while \citet{gnoyke2024evolution} examined architectural-smell evolution patterns. Our results show that, at the code level, the stability of a class is associated not only with code smells in the class itself, but also with smells in the class's efferent neighbors and with interactions between these smells. Future research could examine whether a similar relationship exists at the architectural level. When studying the evolution of a selected architectural smell in a component, researchers could also examine architectural smells in other components on which that component depends, i.e., its efferent neighbor components. They could then investigate whether those architectural smells and their interactions with the selected smell are associated with its evolution characteristics.

\section{Threats to Validity}
\label{sec:threats}

In this section, we present what we believe are the most critical threats.

A threat to internal validity stems from the temporal mismatch between measuring code smells and dependencies at a single snapshot and measuring class stability over a subsequent one-year observation window. Because code smells and dependencies may evolve during the window, predictor values measured at the initial snapshot may differ from those at later points in time. We mitigate this threat by selecting an observation window that is long enough to capture meaningful maintenance activity, but not so extended that predictors are likely to evolve substantially. However, since we cannot fully rule out time-varying changes in predictors within the window, our results should be interpreted as associations between snapshot conditions and subsequent stability rather than evidence of causal effects.

Another threat to internal validity is our choice of a one-year commit history as the observation window, since different time spans may produce different stability patterns. We selected a one-year window rather than the full project lifecycle because our goal is to investigate how code smells in a class's efferent neighbors relate to the class's subsequent stability, rather than to characterize how class stability evolves across the entire project lifecycle. Additionally, as the time span grows longer, it becomes more likely that (1) historical behavior and age-related lifecycle effects are captured (i.e., as a project becomes older, its development and maintenance practices may evolve), (2) a class's functionality changes substantially, and (3) the temporal mismatch discussed in the previous paragraph becomes more severe. These factors can confound the relationship between predictors and the measured stability outcomes. Shorter periods, in contrast, may not include enough maintenance activity to meaningfully assess stability. To mitigate the risk of insufficient maintenance activity, our system selection criteria require a certain level of development activity during the observation window, and we report descriptive statistics (e.g., commits and churn) to empirically characterize the maintenance activity during the one-year period.

Another threat to internal validity is that the studied projects may differ systematically in many factors, such as age and commit practices. These factors may lead to different baseline levels of class stability independent of the studied predictors. To mitigate this risk, we include a project-level random intercept to account for project-specific baseline differences. This adjustment helps ensure that class-level comparisons are made on a more comparable basis across projects, and supports findings that generalize beyond any single project.

A further threat to internal validity arises from our use of stricter detection thresholds to reduce false positives. This choice may increase false negatives. Undetected smell instances may lead to underestimation of smell-based variables, including those measuring code smells in the focal class and its efferent neighbors, as well as efferent code smell coupling and interaction. This measurement error may make the estimated associations between these variables and class stability appear weaker than they actually are.

A threat to external validity concerns our project selection process. Following the recommendations of \citet{kalliamvakou2016depth}, we exclude inactive and educational repositories and focus on widely adopted projects to maximize insights into real code evolution. However, we limit our analysis to Java systems due to the language support of the smell-detection tool. While this procedure reduces noise introduced by low-quality or non-representative repositories, it also limits the extent to which our results generalize to non-Java or less popular projects.

A threat to construct validity is that our classification of a class as non-smelly is limited to the 10 code smell types considered in this study. Consequently, a class classified as non-smelly may still contain other smell types not examined here. However, as discussed in Section~\ref{sec:background}, our goal is not to exhaustively cover all possible code smells, but to focus on a representative subset.

The selection of the code smell detector is a potential threat to construct validity because existing tools differ in their detection rules, documentation, and the range of smells they support. To reduce this threat, we implemented our own detector following the metric-based detection strategies proposed by \citet{lanza2007object}. We make the detector implementation available in our replication package \citep{effint2026zhang} to support transparency and replicability.

Another threat to construct validity is the discrepancy between tool-detected code smells and the design problems perceived by developers. While subjective judgments can never be fully captured, we reduce this discrepancy by using our own tool that closely follows the widely adopted smell definitions of \citet{lanza2007object}. Because these definitions underpin many refactoring guidelines, smell-detection tools, and code smell studies, they approximate a community-level consensus among experienced developers. As a result, the detector's output is more likely to align with their expectations than tools that rely on opaque or non-standard detection rules. Nevertheless, individual developers may still disagree with specific detections depending on domain knowledge, design intent, or business logic, and our findings should be interpreted with this limitation in mind.

Another threat to construct validity is that we treat all dependency types uniformly (i.e., without dependency-type-specific weighting) when defining efferent neighbors and deriving interaction measures, although different dependency types may induce ripple effects with different strengths. This decision is motivated by the lack of prior evidence to justify dependency-type-specific weighting; applying such weighting would introduce subjective design choices. Consequently, our results should be interpreted as aggregate effects across dependency types, rather than dependency-type-specific mechanisms.

A further threat to construct validity is that we do not include \emph{Duplicated Code} (i.e., multiple similar or identical code fragments that appear in different locations \citep{roy2007survey}) as a smell in this study. Our theory focuses on ripple effects that propagate along explicit static dependencies between classes, whereas changes related to duplicated code can spread through clone similarity and clone-management practices (e.g., copy-paste) without explicit dependency links. Consequently, clone-related propagation mechanisms may not be captured by our dependency-based measurements, and our findings should be interpreted as applying to dependency-driven rather than clone-driven change propagation.

A threat to conclusion validity concerns the suitability of our statistical analysis. To verify the distributional assumptions, we fitted Poisson GLMMs and computed dispersion statistics. The diagnostics indicated overdispersion for both \texttt{ChF} and \texttt{ChS}, so we used negative binomial GLMMs, which are well-suited for overdispersed count data and were fitted using the established glmmTMB package in R. Because the model family was selected based on explicit distributional checks and relies on well-validated methods, the risk that our conclusions are driven by the choice of model or software is reduced.

\section{Conclusions} \label{sec:conclusions}

This study examined class stability using change frequency and change size. We investigated whether class stability is associated not only with code smells in the class itself, but also with code smells in its efferent neighbors (i.e., the classes it depends on), efferent code smell coupling, and efferent code smell interaction. We analyzed one year of history from 152 open-source Java projects, covering 287,580 classes and 10 code smell types.

Our results challenge the practice of examining code smells only within the class being maintained (i.e., the focal class) by showing that important stability risks can be missed when the analysis is limited to that class. Code smells are substantially more prevalent in efferent neighbors (77\%) than in focal classes (19\%). Beyond this prevalence difference, considering code smells in efferent neighbors reveals stability risks that are not observed when examining code smells in the focal class alone. After accounting for code smell presence in the focal class and in its efferent neighbors, efferent code smell coupling shows an additional association with 17.8\% higher change frequency and 8.3\% higher change size. Moreover, analyses of specific smell-type pairs show that considering efferent code smell interactions reveals additional stability risks that are not observed when the corresponding smell types are analyzed separately in the focal class or in its efferent neighbors.

Taken together, these findings suggest that code smell analysis for class maintenance should extend beyond the class being maintained to the classes on which it depends. Our findings can therefore inform maintenance prioritization and provide directions for future research on smells and their relationships across dependencies.

Future research could investigate a broader range of code smells and apply a similar study design to smells at other levels of the system, such as architectural smells. It could also investigate whether the association between efferent code smell interactions and lower class stability depends on the dependency type connecting the smells. Another direction is to assess whether the observed stability associations differ across contextual factors such as programming language and project age. Additional maintenance outcomes beyond change frequency and change size could also be considered, such as defect-proneness, to examine whether similar associations are observed for other aspects of software maintainability and quality.

\section*{Declarations} \label{data_avai}

\noindent\textbf{Funding.} This research received no funding.

\vspace{1.0em}
\noindent\textbf{Ethical Approval.} Not applicable.

\vspace{1.0em}
\noindent\textbf{Informed Consent.} Not applicable.

\vspace{1.0em}
\noindent\textbf{Author Contributions.} \textbf{Zushuai Zhang:} Conceptualization, Methodology, Software, Investigation, Data curation, Formal analysis, Validation, Visualization, Writing -- original draft, Writing -- review \& editing. \textbf{Elliott Wen:} Validation, Supervision, Writing -- review \& editing. \textbf{Ewan Tempero:} Validation, Supervision, Writing -- review \& editing.

\vspace{1.0em}
\noindent\textbf{Data Availability Statement.} We have made our source code, scripts, processing pipeline, datasets, and replication instructions available in our replication package \url{https://figshare.com/s/7b0126aa32a04c98abd4} \citep{effint2026zhang}. \textbf{Ethical Considerations:} We acknowledge that GitHub repository data may contain person-related information (e.g., developer identifiers). Therefore, rather than redistributing copies of the GitHub repositories, our replication package \citep{effint2026zhang} provides the GitHub repository link for each analyzed project, a script for downloading all repositories used in this study directly from GitHub, and the start and end commits defining the analyzed history for each repository. Together, these materials enable the repository data used in this study to be reconstructed for replication.

\vspace{1.0em}
\noindent\textbf{Conflict of Interest.} The authors declare that they have no conflict of interest.

\appendix

\section{Complete Primary Analysis Results}
\label{app:primary_results}

Table~\ref{tab:primary_results_full} reports the complete results for all 34 primary analysis tests, including $\beta$ (the coefficient of the IV in the model), its standard error (SE), and 95\% confidence interval (95\% CI). We report both unadjusted and BH-adjusted $p$-values, with significant BH-adjusted values ($p<0.05$) shown in bold.

\begin{table*}[ht]
\centering
\caption{Complete primary-analysis results.}
\label{tab:primary_results_full}
\resizebox{\textwidth}{!}{%
\begin{tabular}{@{}lrrlrrrrr@{}}
\hline
\textbf{Hypothesis} & $\boldsymbol{\beta}$ & \textbf{SE} & \textbf{95\% CI} & \textbf{IRR} & \textbf{AME} & \makecell{\textbf{Doubling}\\\textbf{assoc.}} & \makecell{\textbf{Unadjusted}\\$\boldsymbol{p}$} & \makecell{\textbf{BH}\\$\boldsymbol{p}$} \\
\hline
\multicolumn{9}{l}{\textbf{RQ1}} \\
$\boldsymbol{H1.1^{\text{\texttt{ChF}}}}$ & 0.01132 & 0.00865 & [-0.00564, 0.02828] & 1.01139 & 0.00731 & -- & 0.09532 & 0.19065 \\
$\boldsymbol{H1.1^{\text{\texttt{ChS}}}}$ & -0.01864 & 0.01610 & [-0.05020, 0.01291] & 0.98153 & -0.21325 & -- & 0.87658 & 0.87658 \\
$\boldsymbol{H1.2^{\text{\texttt{ChF}}}}$ & 0.05906 & 0.00738 & [0.04460, 0.07353] & 1.06084 & 0.02702 & 1.04179 & $6.09\times10^{-16}$ & $\boldsymbol{3.65\times10^{-15}}$ \\
$\boldsymbol{H1.2^{\text{\texttt{ChS}}}}$ & 0.00587 & 0.01477 & [-0.02308, 0.03482] & 1.00589 & 0.04274 & 1.00408 & 0.34555 & 0.41466 \\
$\boldsymbol{H1.3^{\text{\texttt{ChF}}}}$ & 0.04486 & 0.00933 & [0.02658, 0.06313] & 1.04588 & 0.02119 & 1.03158 & $7.54\times10^{-7}$ & $\boldsymbol{2.26\times10^{-6}}$ \\
$\boldsymbol{H1.3^{\text{\texttt{ChS}}}}$ & 0.01346 & 0.01806 & [-0.02194, 0.04886] & 1.01355 & 0.10249 & 1.00937 & 0.22802 & 0.34203 \\
\multicolumn{9}{l}{\textbf{RQ2}} \\
$\boldsymbol{H2.1^{\text{\texttt{ChF}}}}$ & -0.05483 & 0.01051 & [-0.07542, -0.03424] & 0.94665 & -0.03603 & -- & 1 & 1 \\
$\boldsymbol{H2.1^{\text{\texttt{ChS}}}}$ & -0.09098 & 0.01667 & [-0.12365, -0.05830] & 0.91304 & -1.08295 & -- & 1 & 1 \\
$\boldsymbol{H2.2^{\text{\texttt{ChF}}}}$ & -0.01456 & 0.00384 & [-0.02209, -0.00703] & 0.98554 & -0.00209 & 0.98996 & 0.99992 & 1 \\
$\boldsymbol{H2.2^{\text{\texttt{ChS}}}}$ & -0.02968 & 0.00651 & [-0.04244, -0.01692] & 0.97076 & -0.07283 & 0.97964 & 1 & 1 \\
$\boldsymbol{H2.3^{\text{\texttt{ChF}}}}$ & 0.02457 & 0.00817 & [0.00856, 0.04057] & 1.02487 & 0.00623 & 1.01717 & 0.00131 & \textbf{0.00526} \\
$\boldsymbol{H2.3^{\text{\texttt{ChS}}}}$ & 0.07487 & 0.01376 & [0.04790, 0.10184] & 1.07775 & 0.31896 & 1.05327 & $2.65\times10^{-8}$ & $\boldsymbol{3.01\times10^{-7}}$ \\
$\boldsymbol{H2.4^{\text{\texttt{ChF}}}}$ & -0.03740 & 0.01204 & [-0.06100, -0.01380] & 0.96329 & -0.01665 & -- & 0.99905 & 1 \\
$\boldsymbol{H2.4^{\text{\texttt{ChS}}}}$ & -0.07384 & 0.01951 & [-0.11208, -0.03560] & 0.92882 & -0.50035 & -- & 0.99992 & 1 \\
$\boldsymbol{H2.5^{\text{\texttt{ChF}}}}$ & -0.01581 & 0.00448 & [-0.02460, -0.00703] & 0.98431 & -0.00228 & 0.98910 & 0.99979 & 1 \\
$\boldsymbol{H2.5^{\text{\texttt{ChS}}}}$ & -0.02245 & 0.00766 & [-0.03747, -0.00743] & 0.97780 & -0.05259 & 0.98456 & 0.99831 & 1 \\
$\boldsymbol{H2.6^{\text{\texttt{ChF}}}}$ & 0.01916 & 0.00981 & [-0.00007, 0.03840] & 1.01935 & 0.00423 & 1.01337 & 0.02544 & 0.07631 \\
$\boldsymbol{H2.6^{\text{\texttt{ChS}}}}$ & 0.08765 & 0.01646 & [0.05540, 0.11991] & 1.09161 & 0.29844 & 1.06264 & $5.02\times10^{-8}$ & $\boldsymbol{3.01\times10^{-7}}$ \\
\multicolumn{9}{l}{\textbf{RQ3}} \\
$\boldsymbol{H3.1^{\text{\texttt{ChF}}}}$ & 0.03095 & 0.00922 & [0.01289, 0.04902] & 1.03144 & 0.02003 & -- & 0.00039 & \textbf{0.00078} \\
$\boldsymbol{H3.1^{\text{\texttt{ChS}}}}$ & -0.00881 & 0.01768 & [-0.04347, 0.02585] & 0.99123 & -0.10072 & -- & 0.6908 & 0.6908 \\
$\boldsymbol{H3.2^{\text{\texttt{ChF}}}}$ & 0.16376 & 0.02433 & [0.11608, 0.21145] & 1.17794 & 0.10762 & -- & $8.41\times10^{-12}$ & $\boldsymbol{3.36\times10^{-11}}$ \\
$\boldsymbol{H3.2^{\text{\texttt{ChS}}}}$ & 0.08005 & 0.03896 & [0.00369, 0.15641] & 1.08334 & 0.92383 & -- & 0.01996 & \textbf{0.02661} \\
\multicolumn{9}{l}{\textbf{RQ4}} \\
$\boldsymbol{H4.1^{\text{\texttt{ChF}}}}$ & 0.01464 & 0.00971 & [-0.00439, 0.03368] & 1.01475 & 0.00946 & -- & 0.06585 & 0.09877 \\
$\boldsymbol{H4.1^{\text{\texttt{ChS}}}}$ & 0.02356 & 0.01894 & [-0.01356, 0.06067] & 1.02384 & 0.27006 & -- & 0.10675 & 0.14233 \\
$\boldsymbol{H4.2^{\text{\texttt{ChF}}}}$ & 0.01647 & 0.00358 & [0.00945, 0.02350] & 1.01661 & 0.00705 & 1.01148 & $2.16\times10^{-6}$ & $\boldsymbol{2.59\times10^{-5}}$ \\
$\boldsymbol{H4.2^{\text{\texttt{ChS}}}}$ & 0.01694 & 0.00728 & [0.00267, 0.03122] & 1.01709 & 0.11286 & 1.01181 & 0.01001 & \textbf{0.02002} \\
$\boldsymbol{H4.3^{\text{\texttt{ChF}}}}$ & 0.00925 & 0.00308 & [0.00322, 0.01528] & 1.00929 & 0.00393 & 1.00643 & 0.00133 & \textbf{0.00398} \\
$\boldsymbol{H4.3^{\text{\texttt{ChS}}}}$ & 0.01490 & 0.00618 & [0.00279, 0.02700] & 1.01501 & 0.09759 & 1.01038 & 0.00792 & \textbf{0.01901} \\
$\boldsymbol{H4.4^{\text{\texttt{ChF}}}}$ & -0.06263 & 0.01322 & [-0.08854, -0.03672] & 0.93929 & -0.04018 & -- & 1 & 1 \\
$\boldsymbol{H4.4^{\text{\texttt{ChS}}}}$ & 0.04987 & 0.02662 & [-0.00231, 0.10205] & 1.05113 & 0.56759 & -- & 0.03053 & 0.05234 \\
$\boldsymbol{H4.5^{\text{\texttt{ChF}}}}$ & -0.00779 & 0.00541 & [-0.01840, 0.00281] & 0.99224 & -0.00332 & 0.99461 & 0.92518 & 1 \\
$\boldsymbol{H4.5^{\text{\texttt{ChS}}}}$ & 0.04151 & 0.01106 & [0.01983, 0.06319] & 1.04239 & 0.27507 & 1.02919 & $8.74\times10^{-5}$ & \textbf{0.00047} \\
$\boldsymbol{H4.6^{\text{\texttt{ChF}}}}$ & -0.01535 & 0.00444 & [-0.02406, -0.00664] & 0.98477 & -0.00649 & 0.98942 & 0.99972 & 1 \\
$\boldsymbol{H4.6^{\text{\texttt{ChS}}}}$ & 0.03342 & 0.00909 & [0.01560, 0.05123] & 1.03398 & 0.21773 & 1.02343 & 0.00012 & \textbf{0.00047} \\
\hline
\end{tabular}%
}
\end{table*}

\clearpage
\bibliography{sections/References}

\end{document}